# Method: Using generalized additive models in the livestock animal sciences

G. L. Simpson[a]

[a] *Aarhus University, Department of Animal and Veterinary Sciences, Blichers Allé 20, Tjele, Denmark, 8830*

**Abstract**

Nonlinear relationships between covariates and a response variable of interest are frequently encountered in animal science research. Within statistical models, these nonlinear effects have, traditionally, been handled using a range of approaches, including transformation of the response, parametric nonlinear models based on theory or phenomenological grounds (e.g., lactation curves), or through fixed degree spline or polynomial terms. If it is desirable to learn the shape of the relationship of one or more covariates and a response, then generalized additive models (GAMs) are an excellent alternative to these traditional approaches. GAMs extend the generalized linear model such that the linear predictor includes one or more smooth functions, parameterised using penalised splines. A wiggliness penalty on each function is used to avoid over fitting while estimating the parameters of the spline basis functions to maximise fit to the data. Modern GAMs include automatic smoothness selection methods to find an optimal balance between fit and complexity of the estimated functions. Because GAMs learn the shapes of functions from the data, the user can avoid forcing a particular model to their data. Here, I provide a brief description of GAMs and visually illustrate how they work. I then demonstrate the utility of GAMs on three example data sets of increasing complexity, to show i) how learning from data can produce a better fit to data than that of parametric models, ii) how hierarchical GAMs can be used to estimate growth data from multiple animals in a single model, and iii) how hierarchical GAMs can be used for formal statistical inference in a designed experiment of the effects of exposure to maternal hormones on subsequent growth in Japanese quail. In each case I show how the fitted model can be used to estimate biologically-motivated values (e.g., persistence of the lactation curve). The examples are supported by R code that demonstrates how to fit each of the models considered, and reproduces the results of the statistical analyses reported here. Ultimately, I show that GAMs are a modern, flexible, and highly usable statistical model that is amenable to many research problems in animal science.

## Implications

Nonlinear relationships between covariates and a response variable of interest are frequently encountered in animal science research. Generalized additive models with automatic smoothness selection via penalised splines provide an attractive, flexible, data-driven statistical model that is capable of estimating these relationships. I provide a description of the generalized additive model and demonstrate its use on three typical data examples; i) a lactation curve from a dairy cow, ii) growth curves in commercial pigs, and iii) an experiment on the effects of maternal hormones on growth rates in Japanese quail.

*Corresponding author
*Email address:* gavin@anivet.au.dk (G. L. Simpson)

**Specification table**

| | |
|---|---|
| Subject | Livestock farming systems |
| Specific subject area | Quantitative analysis of animal performance, growth, and modelling |
| Type of data | Table, graph, code |
| How data were acquired | Lactation curve: unstated in original source.<br>Pig growth data: weight measurements derived from a depth camera (iDOL65, dol-sensors a/s, Aarhus, Denmark) and a YOLO (you only look once) algorithm.<br>Quail growth experiment: body mass recorded using a digital balance. |
| Data format | Lactation curve: processed (averaged).<br>Pig growth data: processed (averages of multiple depth camera-based weight measurements).<br>Quail growth experiment: Raw. |
| Parameters for data collection | Lactation curve: daily fat content of milk for a single animal (cow 7450).<br>Pig growth data: Pigs raised under conventional husbandry conditions at two farms.<br>Quail growth experiment: 158 eggs from adult Japanese quails provided by Finnish private local breeders were injected with one or a combination of maternal hormones or a saline solution control, and those eggs that hatched successfully were monitored for 78 days for a range of parameters including body mass. |
| Description of data collection | Lactation curve: fat content of daily milk production was measured for a single cow.<br>Pig growth data: a depth camera observed the pigs *in situ* and a computer algorithm converted the digital imagery of individual animals into weight estimates.<br>Quail growth experiment: the body mass of hatched quail was recorded at 12 hours post hatching, once every three days for days three to 15, and weekly thereafter until day 78. |
| Data source location | Lactation curve: unknown.<br>Pig growth data: Data from two farms were reported in the original study: A commercial farm in Gronau, Germany, and the experimental farm of the Department of Animal and Veterinary Sciences, Aarhus University, Viborg, Denmark.<br>Quail growth experiment: Finland |
| Data accessibility | Repository name: Zenodo<br>Data identification number: 10.5281/zenodo.15777270<br>Direct URL to data: https://doi.org/10.5281/zenodo.15777270 |

## Introduction

Many research questions in the animal sciences involve the estimation of nonlinear relationships between covariates and a response variable of interest. A classic example, with a long history of statistically-based and mathematically-based research, is the plethora of models that have been described for the estimation of lactation curves from test day data or automatic milking machines (e.g., Macciotta et al., 2011). Another, is the estimation of growth curves in the context of breeding and genetics (e.g., White et al., 1999).

Generalized additive models (GAMs, Hastie and Tibshirani, 1990; Wood, 2017) are a powerful and flexible class of regression models, which extend the generalized linear model (GLM) to allow the effects of covariates on the response to be modelled using smooth, nonlinear functions. Despite the frequency with which such nonlinear relationships are encountered, surprisingly little use of GAMs has been seen in animal science to date. Notable exceptions include Hirst et al. (2002), Yano et al. (2014), Benni et al. (2020), and Huang et al. (2023). The particular advantage of GAMs over parametric models is that because GAMs do not force a particular model on to the data thet can adapt to the data themselves, providing a substantially improved estimation of covariate effects and thence the biological interpretations derived from these.

In part, the lack of uptake of GAMs in animal science reflects a traditional statistics workflow grounded in linear mixed effects modelling. Statistical training rarely includes more advanced models like GAMs, and GAMs are sufficiently different an approach that many researchers may be wary of using them because they are unfamiliar with the nomenclature used to describe the models or the software used to fit them. Where GAMs have been used in animal science, best practice is often not followed, for example in the choice of smoothness selection method, or failing to correctly specify the conditional distribution of the response (e.g. van Lingen et al., 2023).

More attention has been paid to the use of spline terms within linear models, frequently in comparisons against some form of polynomial-based model, especially Legendre polynomials (e.g., Nagel-Alne et al., 2014; Silvestre et al., 2006; Macciotta et al., 2010; Brito et al., 2017). There, the focus has largely been on choosing the number of knots in the spline basis expansion and on the placement of those knots. Modern GAMs largely make such choices redundant; with penalised splines, a wiggliness penalty is used to avoid over-fitting, and low-rank eigen bases, such as the low-rank thin plate regression spline basis of Wood (2003), avoid the knot-placement issue for most problems.

A guide to the use of GAMs in the animal science setting, written with users in mind, is needed to raise awareness of the utility of these flexible models and to promote best practice in their fitting. Below, I describe GAMs and demonstrate visually how they work. Then I apply GAMs to three different examples that scale in complexity and which represent typical data problems encountered in animal science. As well as showing the predictive ability of GAMs, the examples demonstrate how the fitted models can be used to provide estimates that are biologically relevant, such as treatment differences, estimates of growth rates, or persistency in milk production. The examples are supported by tutorials that contain the computer code needed to fit the models in the R statistical software (R Core Team, 2025), which are available alongside the source code for the manuscript itself.

## Materials and methods

I begin with an introduction to GAMs and how they work. This section is necessarily general to allow for a brief summary of the main points and illustration of how penalised splines work. In the subsequent section describing the three worked examples I provide more case-specific descriptions of the models that are fitted to data.

*Generalized additive models*

A basic GAM (Hastie and Tibshirani, 1990) has the following form

$$\begin{aligned} y_i &\sim \mathcal{D}(\mu_i, \phi) \\ \mathbb{E}(y_i) &= \mu_i \\ g(\mu_i) &= \mathbf{X}_i \boldsymbol{\gamma} + \sum_j f_j(\bullet), \; i = 1, \ldots, n; \; j = 1, \ldots, J \end{aligned}$$

where $y_i$ is the $i$th observation of a univariate response variable of interest that is modelled as a function of covariates on the scale of a link function $g()$. $\mathcal{D}(\mu_i, \phi)$ is a distribution, typically from, though not limited to, the exponential family of distributions, with mean $\mu_i$ and scale parameter $\phi$. $\mathbf{X}_i$ is the $i$th row of the model matrix of any parametric terms (including the model intercept or constant term), and $\boldsymbol{\gamma}$ the associated regression parameters. The $f_j$ are $J$ smooth functions of one or more covariates (independent variables, e.g. days in milk); I use $\bullet$ as a placeholder for covariate(s), but in the simplest case of a smooth of a single continuous covariate we have $f_j(x_{ij})$, a univariate smooth where $x_{ij}$ is the $i$th observation of the $j$th covariate. For a smooth of two or more continuous variables, a tensor product smooth representing smooth main effects and smooth interaction, the linear predictor would include $f(x_{i1}, x_{i2})$, for example.

In the remainder of this section, I aim to present *just enough* detail about GAMs to afford the reader a general understanding of what is involved in estimating such a model so that they can appreciate how GAMs work, and how we aim to avoid overfitting or having to choose how complex each smooth function should be. The original approach (Hastie and Tibshirani, 1990) for fitting GAMs, known as *backfitting*, required the analyst to specify how many degrees of freedom each function in the model should take *a priori*; this was viewed by many as being too subjective for more than exploratory analysis. With modern automatic smoothness selection methods this problem has largely been resolved through the use of penalised splines and fast algorithms for smoothness selection (e.g., Wood, 2011; Wood et al., 2016).

*Penalised splines*

In a GAM, the smooth functions $f_j()$ are typically represented in the model using splines, although other functions fit into this framework, most notably iid Gaussian random effects. A spline is composed of $K$ basis functions, $b_k()$, and their associated coefficients, $\beta_k$

$$f_j(x_{ij}) = \sum_{k=1}^{K} \beta_k b_k(x_{ij}) \tag{1}$$

where the basis functions $b_k()$ could be piecewise cubic polynomials for a cubic regression spline, but more generally are just the component functions of a spline (e.g. B splines, thin plate splines). For identifiability reasons, the basis is subject to a sum-to-zero constraint to remove the constant function from the span of the basis, which allows inclusion of a separate constant term or intercept in the model; this is desirable if we also want to include categorical (factor) terms in the model for example.

Figure 1a shows a cubic B spline ($K = 10$ basis functions) fitted to simulated data. Fitting a spline involves finding estimates for the coefficients of basis functions, $\beta_k$. These coefficients weight the individual basis functions as shown in Figure 1a. To find the value of the fitted spline at each value of $x$, we sum up the values of the weighted basis functions evaluated at each value of $x$. This yields the light blue curve in Figure 1a. The coefficients for the basis functions, $\beta_k$, are determined by forcing the fitted function to go as close to the data as possible. As shown in Figure 1a, we largely recover the true function from which the data were simulated. The B spline basis functions for a covariate $x$ after the absorbing the sum-to-zero constraint into the basis are shown in Figure 1b.

Expanding $\mathbf{x}_j$ into many basis functions in this way means we must be cognizant of the risk of over fitting our sample of data. Taken to the extreme, we could obtain an arbitrarily close fit to the data by using as many basis functions as there are data (i.e., $K = n$), but all this would achieve in practice is the replacement of the data with a set of coefficients.

This raises the question of how many basis functions should be used? If there were $n$ unique (i.e., no duplicated values) values of the covariate $x$ in the data set, one option would be to use $n$ basis functions, leading to a situation where our model would have as many coefficients as unique data. We gain very little however by using $n$ basis functions from a statistical viewpoint. In practice, we typically use $\mathcal{N} \ll n$ basis functions. Yet, even with just $\mathcal{N}$ basis functions, we face the real possibility that our model may overfit the data if $\mathcal{N}$ is too large. To address this, modern approaches to fitting GAMs use penalised splines with automatic smoothness selection (e.g. Eilers and Marx, 1996; Wood, 2011; Wood et al., 2016).

If our aim is to avoid overfitting, we need to penalise highly complex fitted functions, and hence need to define what we mean by "complex". A complex function would be one that "wiggles" about markedly as we move from low to high values of the covariate, $x$. Such wiggling around implies that the function has a high amount of curvature, which we measure using the second derivative of the fitted function. While there are other definitions for complexity that we can use, this is the typical measure used for penalized splines in statistical models. Therefore, to measure the wiggliness of a fitted function we need to integrate (sum up) the second derivative of the function $f$ over the range of $x$

$$\int_x f''(x)^2 \, dx = \boldsymbol{\beta}^\mathsf{T} \mathbf{S} \boldsymbol{\beta} \tag{2}$$

where $f''$ indicates the second derivative of $f$, and note that we are integrating the *square* of the second derivative because we need to allow for both negative and positive curvature as the function wiggles. Conveniently, we can compute this integral as a function of the model coefficients $\boldsymbol{\beta}$ as shown on the right hand side of Equation 2. $\mathbf{S}$ is a known penalty matrix, which encodes the complexity of the basis functions. The penalty matrix for the basis expansion shown in Figure 1a is displayed in Figure 1c. Typically, the software used to fit the GAM, such as the R package *mgcv* that is used here, will calculate $\mathbf{S}$ for each spline in the model.

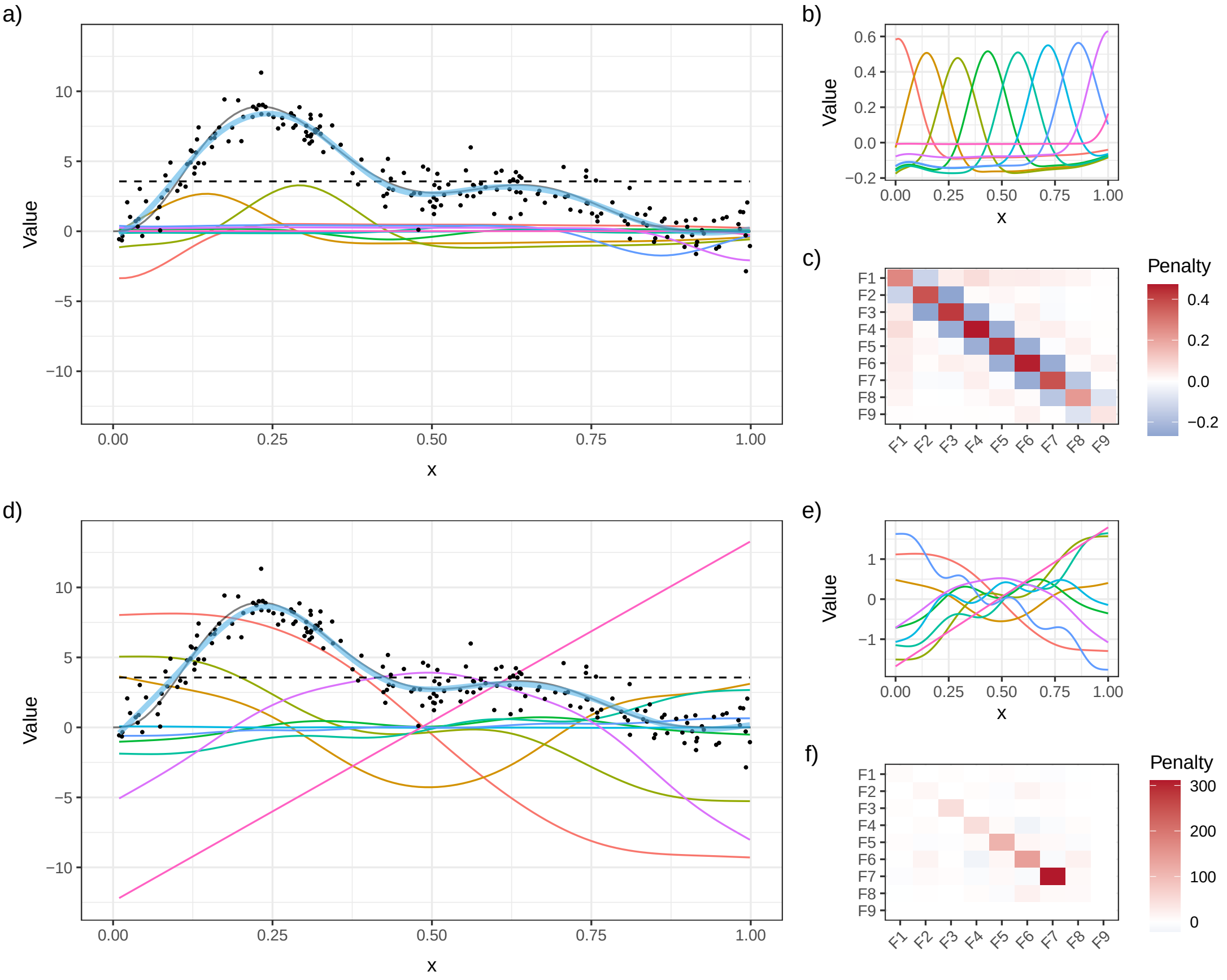

Fig. 1: Illustration of how penalised splines work. A spline basis expansion (a) and associated penalty matrix $\mathbf{S}$ (c) are formed for a covariate $x$. Model fitting involves finding estimates for the coefficients of the basis functions that make the fitted spline (thick, blue curve) go as close to the data (black points) as possible, without over fitting (a). In (a) and (b) the basis functions are shown as thin coloured lines and are from a cubic B spline basis. The thin black line in (a) is the true function from which the data were simulated. The dashed horizontal line in (a) is the estimated value of the intercept. The sum-to-zero identifiability constraint needed so that an intercept can be included in the model has been absorbed into the basis shown in (a) and (b). The penalty matrix (c) encodes how wiggly each basis function is in terms of its second derivative. Panels (d) – (e) are as above but for a low-rank thin plate spline basis.

Figure 2 shows three different estimated smooths for the simulated data shown in Figure 1, plus their wiggliness or integrated squared second derivative. Fitting a GAM requires us to balance the fit to the data *and* the complexity of the resulting model; we wish to avoid both over fitting the data (Figure 2a) and over smoothing (Figure 2c) the data. We want the fitted functions to be just "wiggly enough" to approximate the true, but unknown functions (Figure 2b).

To that end, we find estimates of the basis function coefficients, $\beta_k$, that make the fitted spline go as close to the data as possible, without overfitting; in practice, this is done by maximising the penalised log likelihood

$$\ell_p(\boldsymbol{\beta}, \boldsymbol{\lambda}) = \ell(\boldsymbol{\beta}) - \frac{1}{2\phi} \sum_j \lambda_j \boldsymbol{\beta}_j^{\mathsf{T}} \mathbf{S}_j \boldsymbol{\beta}_j \ ,$$

where $\ell_p(\boldsymbol{\beta})$ is the penalised log likelihood and $\ell(\boldsymbol{\beta})$ the log likelihood of the data, given $\boldsymbol{\beta}$, and the remainder is the wiggliness penalty for the model. The log likelihood ($\ell(\boldsymbol{\beta})$) measures how well our model fits the data, while the wiggliness penalty $\boldsymbol{\beta}_j^\mathsf{T}\mathbf{S}_j\boldsymbol{\beta}_j$ measures how complex the model is. Note that any parametric coefficients, $\boldsymbol{\gamma}$, including the intercept have been absorbed into $\boldsymbol{\beta}$ for convenience. The $\lambda_j$ are known as the smoothing parameters of the model, and it is these smoothing parameters that actually control how much the wiggliness penalty affects the $\ell_p(\boldsymbol{\beta})$. We can think of the $\lambda_j$ as tuning or hyper-parameters of the model.

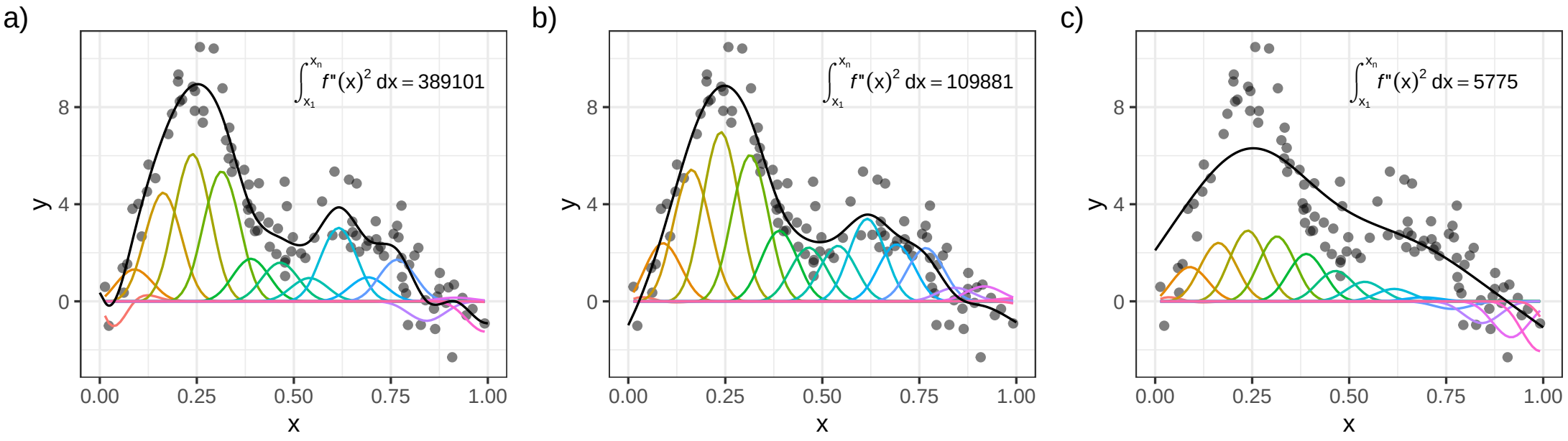

Fig. 2: Illustration of how the wiggliness penalty controls the resulting fit of a penalised spline. The weighted basis functions are shown as thin coloured lines. In each panel a penalised spline is shown by the solid black line, which has been fitted to the data points shown. The wiggliness value of the spline, the integrated squared derivative of the fitted spline over $x$ is given in the upper right of each panel. The spline in (a) is over fitted to the data, resulting in a very wiggly function with a large wiggliness value. The spline in (c) is over smoothed, resulting in a simple fitted function with low wiggliness, but which does not fit the data well. The spline in (b) represents a balance between fit to the data and complexity of fitted function. The smoothing parameter for the spline, $\lambda$, is used as a tuning parameter in the model, which ultimately controls this balance between fit and complexity.

As mentioned in the introduction, the knot placement problem can largely be averted through the use of a low rank thin plate regression spline basis (Figure 1 panels d–f). The spline basis described previously was a cubic B spline basis, and for that basis the knots were placed at evenly-spaced *quantiles* of the covariate. In general, the exact positioning of the knots is unimportant, so long as they are spread out over the range of the covariate. If we wish to avoid the knot placement altogether, we could replace the B spline basis with a low rank thin plate regression spline basis (TPRS, Wood, 2003). We start with a basis function at each unique value of the covariate and then transform and truncate the entire basis with an eigen decomposition and retain the $K$ eigenvectors with the smallest eigenvalues as $K$ new basis functions. The decomposition removes much of the excessive wiggliness that $n$ basis functions would provide, while retaining many of the good properties of the original basis (Wood, 2003). The main downside of the TPRS basis is that it is computationally expensive to form when setting up the model; for large data problems, with many thousands of data, simpler bases such as the cubic regression spline or B spline may be preferred. In the *mgcv* software used here, the TPRS basis is the default basis used for univariate smooths (Wood, 2025, 2011).

The algorithms used to fit GAMs need to estimate the model coefficients, $\boldsymbol{\beta}$, and choose appropriate values of the smoothing parameters, $\lambda_j$. This process is known as smoothness selection, and there are several approaches to smoothness selection that can be taken. One is to treat the problem as one of prediction, and choose $\lambda_j$ in such a way as to minimise the cross-validated prediction error of the model. In practice it would be computationally costly to actually cross-validate the model for fitting, so we approximate the prediction error by minimising the generalised cross validation (GCV) error, AIC, or similar measure. An alternative means of smoothness selection is to take a Bayesian view of the smoothing process (see Miller, 2025, for an accessible introduction to this viewpoint); in doing so, the $\boldsymbol{\beta}_j^{\mathsf{T}}\mathbf{S}_j\boldsymbol{\beta}_j$ can be viewed as multivariate normal priors on the $\boldsymbol{\beta}$. From this Bayesian view of smoothing, we find that the criterion we wish to minimise is that of a mixed effects model. Hence, we can think of the wiggly parts of the $f_j$ as being fancy random effects, while the smooth parts of the $f_j$ are fixed effects, and the $\lambda_j$ are inversely proportional to the random effect variances one would observe if the model were fitted as a mixed effects model. The Bayesian approach to smoothing can involve fully Bayesian estimation using Markov chain Monte Carlo (MCMC) or simulation free estimation via the integrated nested Laplace approximation (INLA), or, using the equivalence of splines and random effects, we can take an empirical Bayesian approach, which yields the posterior modes of the $\boldsymbol{\beta}_j$, or the maximum a posteriori (MAP) estimates.

*Why not polynomials?*

At this stage, it is not unreasonable to question whether we need anything so complicated as a penalized spline; would polynomial terms not work just as well? A polynomial model includes powers of $\mathbf{x}$ in the linear predictor, e.g., $\eta_i = \beta_0 + \beta_1 x_i + \beta_2 x_i^2 + \beta_3 x_i^3 + \cdots + \beta_p x_i^p$. This is also a basis expansion of $\mathbf{x}$, where the basis functions are the zeroth, first, second, third etc. powers of $\mathbf{x}$; $b_1(x_i) = x_i^0 = 1$, $b_2(x_i) = x_i^1 = x_i$, $b_3(x_i) = x_i^2$, $b_4(x_i) = x_i^3$, etc. From the point of view of estimating an unknown function, using this polynomial basis expansion poses two significant problems.

First, while Taylor's theorem suggests that polynomials are useful in approximating an unknown function *at a specific point* (e.g., value of $x_i$), this doesn't imply that polynomials will be similarly useful when interest is on the entire domain of $\mathbf{x}$. If we consider the simpler problem of interpolating data, polynomials tend to perform poorly when approximating functions while requiring the curve to pass through the data points and maintain continuous derivatives with respect to $\mathbf{x}$; to meet these requirements the fitted polynomial curve will tend to oscillate wildly, especially toward the edges of the covariate space (e.g. Wood, 2017, §4.2, pp. 162–4). This problem persists in the regression model setting where we are interested in smoothing data contaminated with noise.

Second, while estimating the polynomial model is trivial, model selection becomes an issue. Model selection is needed to determine the order of polynomial term used in the model as this will ultimately determine the wiggliness of the fitted function. This could be achieved by use of backward selection, while penalized regression with ridge or lasso penalties would shrink specific coefficients thus reducing model complexity. Performing statistical inference (forming confidence intervals, computing $p$ values, etc.) after conducting these model selection procedures is difficult and an active area of research in statistics. Ultimately, this means such approaches tend to be used where the predictive ability of the model is the primary focus. The theory of GAMs is sufficiently advanced that statistical inference is possible currently.

A further problem with polynomial bases is that they are non-local; the fitted value of $y$ at some value $x = x_0$ is very sensitive to the values of $x$ a long way away from $x_0$, especially where the order of polynomial $p$ is greater than 2 or 3.

*Examples*

In the remainder of the section I describe three representative examples that demonstrate the utility of GAMs to address problems in animal science.

*Lactation curves*

As a simple illustration of the benefits of GAMs to learn the functional form of a relationship between a response variable and a covariate, rather than impose one through a parametric model, I reanalyse a small data set of average daily fat content per week of milk from a single cow (Figure 3). The data were reported in Henderson and McCulloch (1990). Initially, I followed their (Henderson and McCulloch, 1990) analysis and fitted a gamma generalized linear model (GLM) with log link function. However, subsequent analysis of this model (and the GAM alternative described below) showed that the average daily fat data are under dispersed relative to that assumed by the gamma distribution. Instead, a Tweedie GLM (log link) was fitted, which has the same response shape as the gamma GLM. The Tweedie distribution is a family of distributions with support on the non-negative real numbers (Smyth, 1996), making it especially suitable for positive or non-negative continuous random variables, such as milk yield, or milk fat content as observed in this example. The Tweedie distribution contains the gamma distribution as a special case, but has more flexibility in terms of the mean-variance relationship assumed of the response; the gamma has a quadratic mean-variance relationship, while those admitted by the Tweedie distributions considered here cover the range from linear (i.e. Poisson-like) to quadratic (i.e., gamma-like) mean-variance scaling.

The Tweedie GLM fitted had the following form

$$\begin{aligned} \texttt{fat}_i &\sim \mathcal{T}(\mu_i, \phi) \\ \log(\mu_i) &= \beta_0 + \beta_1 \log(\texttt{week}_i) + \beta_2 \texttt{week}_i \ . \end{aligned}$$

Henderson and McCulloch (1990) compared the fit of the GLM model with several other formulations and models, including the model of Wood (1967)

$$\texttt{fat}_i = \alpha \, \texttt{week}_i^{\delta} \exp(\kappa \, \texttt{week}_i) + \varepsilon_i$$

where $\alpha$, $\delta$, and $\kappa$ are parameters whose values are to be estimated, and $\varepsilon$ is a Gaussian error term. Note that $\kappa$ is usually presented as $-\kappa$, hence the parameter estimate takes values that are the negative of the usual values. The linear predictors in the GLM and Wood's model are equivalent because of the log link function; $\beta_0 = \log \alpha$, $\beta_1 = \delta$, and $\beta_2 = \kappa$. However, we would not expect both models to produce the same fitted lactation curve as different distributional assumptions are being made; in the GLM version we assume the average daily fat values are conditionally Tweedie distributed, while in Wood's model they are assumed to be conditionally Gaussian distributed.

These models were compared with a GAM version of the log-link, Tweedie GLM, where the fixed functional form of the lactation curve has been replaced by a smooth function of $\texttt{week}_i$, $f(\texttt{week}_i)$, that is to be estimated from the data

$$\begin{aligned} \texttt{fat}_i &\sim \mathcal{T}(\mu_i, \phi) \\ \log(\mu_i) &= \beta_0 + f(\texttt{week}_i) \end{aligned}$$

Wood's (1967) model was estimated via nonlinear least squares using the `nls()` function in R (version 4.5.0, R Core Team, 2025). The Tweedie GLM and GAM were fitted using the `gam()` function of the *mgcv* package (version 1.9.3, Wood, 2025, 2011) for R. $K = 9$ basis functions were used for $f(\texttt{week}_i)$ after application of the identifiability constraint.

The fitted models were subsequently used to estimate some biologically-relevant parameters that could be used to inform breeding or optimise production methods. The week of peak milk fat content, the estimated fat content at the peak, and the rate of decline in the lactation curve at a point midway between the peak and the end of lactation are estimated for all models. For Wood's model, the timing of peak fat content is given by $\delta/\text{-}\kappa$ and peak fat content by $\alpha(\delta/\text{-}\kappa)^{\delta}e^{-\delta}$ (Wood, 1967).

For the Tweedie GAM, we do not have such directly interpretable parameters, therefore the peak is identified by evaluating the model at a grid of 1000 values over the $\texttt{week}_i$ covariate and noting the point at which the estimated fat content is maximised. This process also yields the estimated maximum fat content. While the Tweedie GLM has the same interpretable parameters as Wood's model, determining the uncertainty in the estimate time to peak fat content and peak fat content is not trivial. Instead, the same procedure as that outlined for the GAM was also followed for the GLM.

The first derivative (slope) of the falling limb of a lactation curve is often used as a measure of persistency in dairy animals. While there are several such derivatives that can be derived directly from the parameters of Wood's model (e.g., Dijkstra et al., 2010; Oliveira et al., 2020), exact equivalents for the GAM may not be simple to implement. Instead, for illustration and to avoid differences between models due to the specific details of implementation, for all three models I estimate the first derivative of the lactation curve at a point midway between the peak and the end of lactation. This is intentionally similar to the relative decline measure of Dijkstra et al. (2010). For all three models the first derivative is calculated using a central finite difference with $h$, the step size, set to be 1 day. The central finite difference estimate of the first derivative of the lactation curve, $\hat{f}'$, is

$$\hat{f}' = \lim_{h \to 0} \frac{f\left(x + \frac{h}{2}\right) - f\left(x - \frac{h}{2}\right)}{h} .$$

The uncertainty in each of the three quantities estimated from the models was computed using 10,000 parametric bootstrap samples in the case of Wood's model using the `nlsBoot()` function from package *nlstools* (version 2.1.0, Baty et al., 2015), while for the GLM and GAM 10,000 draws from the posterior distribution of the parameters $\boldsymbol{\beta}$ using a Gaussian approximation to the posterior were used (See Miller, 2025, for a brief introduction to this idea). Posterior sampling was per performed using the `fitted_samples()` function from package *gratia* (Simpson, 2024).

*Pig growth*

In this second example, I illustrate how individual growth curves can be estimated using various forms of hierarchical GAM. A hierarchical GAM (HGAM, Pedersen et al., 2019) is a GAM that includes smooths at different hierarchical levels of the data structure; HGAMs are the GAM equivalent of hierarchical or mixed effects models, which are frequently used with such data. In this example, different parametrisations of the model lead to either direct estimation of each individual animal's growth curve, or a decomposition into an average growth curve plus animal-specific deviations from this average curve. Although previously covered by Pedersen et al. (2019) in detail, a new addition here is the use of a constrained factor smooth to produce animal-specific deviations that are truly orthogonal to the average curve. This basis type was not available to Pedersen et al. (2019), and helps avoid switching to a first derivative penalty for the animal-specific curves that would be required to make the model identifiable if a "factor by" smooth was used.

Automated estimation of body weights is a useful on-farm method for continuously monitoring growth of commercial pigs. I analyse a subset of the weight data reported by Franchi et al. (2023) from a study by Bus et al. (2025), using data from a single pen of 18 pigs (Figure 5). The body weight data were obtained using a depth camera (iDOL65, dol-sensors A/S, Aarhus, Denmark), and the each weight observation is the daily average of multiple measurements made by the camera. As the number of weight measurements made each day varied per animal and per day, each weight observation in the data is the average of a variable number of measurements. To accomodata this, the model included the number of measurements averages as an observation weight, where the precision of each response data is proportional to the number of measurements averaged.

The data exhibit both common (shared) and animal-specific variation. To model these features, four different HGAMs were fitted to the pig weight data. All of the models ultimately provide animal-specific estimates of weight over time, but they decompose the growth curves in different ways. The weight data were assumed to be conditionally gamma distributed, with log link, $\texttt{weight}_i \sim \mathcal{G}(\mu_i, \phi)$, with $\log(\mu_i) = \eta_i$, where $\boldsymbol{\eta}$ is the linear predictor. The gamma distribution was chosen as it has support on the positive real numbers; the weight of a pig is a necessarily positive continuous random variable. Other choices for the conditional distribution would be a log-normal or a Tweedie distribution, but standard model diagnostics did not indicate any problems with the use of the gamma distribution.

The linear predictors for each of the four models were:

$$
\begin{aligned}
&\text{P1: } \eta_i = \beta_{a(i)} + f_{a(i)}(\texttt{day}_i) \\
&\text{P2: } \eta_i = \beta_0 + f_1(\texttt{day}_i) + f_{a(i)}(\texttt{day}_i) \\
&\text{P3: } \eta_i = \beta_0 + f^*_{a(i)}(\texttt{day}_i) \\
&\text{P4: } \eta_i = \beta_0 + f_1(\texttt{day}_i) + f^*_{a(i)}(\texttt{day}_i)
\end{aligned}
$$

where $a(i)$ is an indicator function that gives the animal to which the $i$th observation belongs. Model P1 includes the mean weight of each animal through a parametric factor term, $\beta_{a(i)}$, plus a smooth of observation day $\texttt{day}_i$ *per* animal. The parametric factor term is required because the animal-specific smooths in this model are each subject to the sum-to-zero-constraint and as such do not contain the constant functions that are needed to model the average weight of each animal. These smooths are known informally as "factor by smooths". Model P2 decomposes the data into an average growth curve, $f_1(\texttt{day}_i)$, plus smooths, one per animal, that represent deviations from the average smooth, $f_{a(i)}(\texttt{day}_i)$. Unlike the "factor by smooths", these deviation smooths include constant terms for each animal's average weight, hence only a constant term, $\beta_0$ is include in the parametric part of this model. In both models P1 and P2, the $f_{a(i)}(\texttt{day}_i)$ smooths each have their own smoothing parameters ($\lambda$s), allowing the wiggliness of each animal's growth curve to vary, if supported by the data.

Model P3 is similar to model P1, except that each animal's growth curve, $f^*_{a(i)}(\texttt{day}_i)$, shares a single smoothing parameter ($\lambda$) for the wiggliness and hence assumes that the wiggliness of each curve is similar.

These smooths are denoted with a superscript $*$ to indicate the shared smoothing parameter, and can be thought of as the smooth equivalent of random slopes and intercepts; informally, we refer to these as "random smooths". These smooths are fully penalised and contain constant terms to model the average weight of each animal, hence only the intercept, $\beta_0$ is included in the parametric part of the model. Model P4 is similar to model P2 in terms of the decomposition into an *average* smooth and animal-specific smooth deviations, but, like model P3, the animal-specific smooths share a smoothing parameter and therefore assumes they have similar wiggliness.

The range of GAMs fitted to the pig growth data is intended to illustrate the modelling choices open to the analyst; how are the individual growth curves modelled, and how much variation in wiggliness is allowed per growth curve. The choices then are; is each animal's growth curve modelled directly (P1 and P3) or are the growth curves modelled as an "average" growth curve plus animal-specific deviations from this (models P2 and P4), and should the animal-specific curves be allowed to have different wiggliness if needed (models P1 and P2), or should they all have the same wiggliness (models P3 and P4).

For further details on the different approaches used in the models described above, see Pedersen et al. (2019). Each of the smooths in the models used $K = 9$ basis functions, after application of identifiability constraints.

The estimated daily growth rates on November 15$^{\text{th}}$, 2021 for each pig were calculated from the selected GAM using a central finite difference estimate of the first derivative of the fitted growth curve. The estimated growth rate is given by the median of the posterior distribution and an 95% Bayesian credible interval on the estimate is provided. This date was chosen as weight observations for three of the pigs were missing after November 1$^{\text{st}}$. All estimates were based on 10,000 draws from the posterior distribution, using a Gaussian approximation to the posterior implemented in the functions `response_derivatives()` and `derivative_samples()` from package *gratia*.

*Japanese quail*

In the third example, I demonstrate how to fit GAMs in the context of a designed experiment to obtain estimated treatment effects.

The data (Sarraude et al., 2020a) are from an experiment into the short- and long-term effects on quail of elevated exposure to different types of maternal thyroid hormone in Japanese quail, *Coturnix japonica* (Sarraude et al., 2020b). Briefly, the yolks of $n = 57$ eggs were experimentally manipulated by injection with the prohormone, $T_4$, its active metabolite, $T_3$, or both $T_4$ and $T_3$ ($T_3T_4$), or a saline solution that acted as a control (CO). Body mass was initially measured 12 hours after hatching. Between days 3–15, body mass was recorded every three days, then, between days 15–78, once per week, using a digital balance.

For the same reasons as in the pig growth example, the quail weight data are necessarily positive and were provisionally assumed to be conditionally gamma distributed. Despite this being a reasonable working assumption, model diagnostics identified deviations from this assumption, and instead the data were modelled as being conditionally Tweedie distributed. The Tweedie family of distributions contains the Poisson (power, $p = 1$) and gamma distributions ($p = 2$) as special cases, but is more flexible than gamma, allowing for a range of mean-variance relationships through the power parameter, $p$, of the distribution. In the Tweedie GAMs that were fitted, the power parameter $p$ was allowed to vary between 1–2 and was estimated as an additional model constant term. Models fitted were:

$$\begin{aligned}
&\text{Q1: } \eta_i = \beta_0 + f(\texttt{day}_i) + f_{\text{sex}(i)}(\texttt{day}_i) + f^*_{\text{egg}(i)}(\texttt{day}_i) + \xi_{\text{mother}(i)}\\
&\text{Q2: } \eta_i = \beta_0 + f(\texttt{day}_i) + f_{\text{treat}(i)}(\texttt{day}_i) + f_{\text{sex}(i)}(\texttt{day}_i) + f^*_{\text{egg}(i)}(\texttt{day}_i) + \xi_{\text{mother}(i)}\\
&\text{Q3: } \eta_i = \beta_0 + f(\texttt{day}_i) + f_{\text{treat}(i)}(\texttt{day}_i) + f_{\text{sex}(i)}(\texttt{day}_i) + f_{\text{treat}(i),\text{sex}(i)}(\texttt{day}_i)\\
&\qquad + f^*_{\text{egg}(i)}(\texttt{day}_i) + \xi_{\text{mother}(i)}\\
&\text{Q4: } \eta_i = \beta_0 + f(\texttt{day}_i) + f_{\text{treat}(i)}(\texttt{day}_i) + f_{\text{sex}(i)}(\texttt{day}_i) + f_{\text{treat}(i),\text{sex}(i)}(\texttt{day}_i)\\
&\qquad + \psi_{\text{egg}(i)} + \xi_{\text{mother}(i)}
\end{aligned}$$

where $\texttt{day}_i$ is the number of days since hatching, treat($i$), sex($i$), egg($i$), and mother($i$) indicate to which treatment group, sex, bird, and mother the $i$th observation belongs. Smooth functions with subscript treat($i$), sex($i$), or egg($i$) are factor-smooth interactions representing deviations from the "average" smooth, $f(\texttt{day}_i)$, for the indicated factor. $f_{\text{treat}(i),\text{sex}(i)}(\texttt{day}_i)$ represents a higher order factor-smooth interaction, which allows the time varying treatment effects to also vary between male or female quail. $\psi_{\text{egg}(i)}$ and $\xi_{\text{mother}(i)}$ are iid Gaussian random intercepts for individual quail and their mother respectively. A $f^*$ represents a random smooth, where each smooth in the set shares a smoothing parameter. The factor-smooth interactions in this model were fitted using the constrained factor-smooth interaction basis in the *mgcv* package; this basis excludes the main effects (and lower-order terms in the case of higher-order interactions) to insure that the basis functions are orthogonal to those lower order terms.

Model Q1 represents a null model containing no treatment effects but models the remaining features of the data, decomposing the growth curves into an average effect, a sex-specific effect, and individual quail-specific effects, plus a maternal effect. Model Q2 extends Q1 by adding the treatment effect, $f_{\text{treat}(i)}(\texttt{day}_i)$, which models deviations from the average curve for each of the four treatment levels. Model Q3 extends Q2 to allow different treatment-specific curves for male and female quail. Model Q4 is a variant of Q3, replacing the quail-specific growth curve deviations with an individual level random intercept. *A priori*, model Q3 represents the complete set of hypotheses an analyst might expect to consider for these data.

Each smooth in the quail models used $K = 9$ basis functions after application of identifiability constraints, except the quail-specific smooths, $f^*_{\text{egg}(i)}(\texttt{day}_i)$, which used $K = 6$ basis functions per individual. This reduced number of basis functions per smooth reflects the fact that after accounting for the average shape of the growth curves, plus sex and treatment deviations, any remaining individual-specific variation from these other effects would be smaller in magnitude and less complex (wiggly).

These data were previously analysed using GAMs by Sarraude et al. (2020a) and Morota et al. (2021); Sarraude et al. (2020a) and Morota et al. (2021) use an ARMA(1,1) correlation for model residuals to account for the longitudinal nature of the data that remain unmodelled after inclusion of smooths of bird age conditioned on treatment or sex, and it is unclear if a non-Gaussian distribution was used. The main advance provided by this reanalysis is that instead of treating the individual quail growth curves as stochastic components of the model, here I directly model those curves using "random smooths". Additionally, here I assume the response data are conditionally distributed Tweedie, and I directly assess the treatment by sex interaction.

*Smoothness selection & inference*

Restricted marginal likelihood (REML) smoothness selection (Wood, 2011) was used to estimate each of the GAMs fitted to the example data sets. REML was used because it has slightly better performance in terms of estimating smoothing parameters compared to marginal likelihood (ML) smoothness selection. We did not use GCV smoothness selection for these examples because i) prediction error is a less important consideration here where one is interested in estimation of effects and statistical inference, and ii) GCV is prone to under smoothing (Reiss and Ogden, 2009; Wood, 2011).

In the lactation curve example I use standard model diagnostic techniques (e.g., quantile-qauntile (QQ) plots, plots of residuals versus linear predictor values) to identify the most useful model. In the pig growth example, AIC was used to identify which of the decompositions of time provided the best fit, but the specific choice of model was made using domain knowledge.

In the quail hormone example, AIC values are reported, but *a priori* I chose to report results from model Q3, the "full" model from the point of view of the potential hypotheses under consideration; as the treatment effect was allowed to vary between sexes, I take an estimation-based approach and quantify the magnitudes of any treatment by sex interactions using the most complex model. The remaining models are fitted and reported on briefly for illustrative purposes; performing model term selection (for example, testing the higher order $f_{\text{treat}(i)}(\texttt{day}_i)$ interaction, and deciding on the basis of a $p$ value or AIC whether to retain it in the model or not), would introduce biases into the inference process. As we currently lack post selection inference methods to deal with this model form of selection it is prudent to simply not enter into such model selection procedures.

Using model Q3 allows there to be treatment differences between male and female quail; subsequent estimation of values of interest and comparison among these estimates is instead the inference route followed for the quail example. To illustrate, I estimate two quantities of interest: i) the estimated growth rate (slope) of an average quail at `day` = 20 in all combinations of treatment and sex, and ii) the estimated weight of an average quail at the end of the experiment (`day` = 78), again for all combination of treatment and sex. All pairwise comparisons among treatment levels within sex were performed, with adjustment of $p$ values to control the false discovery rate using the Benjamini–Yekutieli (Benjamini and Yekutieli, 2001) procedure. Estimates and pairwise comparisons were computed using the `slopes()` and `predictions()` functions of the *marginaleffects* package for R (version 0.27.0, Arel-Bundock et al., 2024). Estimates of the expected growth curves for average quail in all combinations of treatment and sex were produced using the `conditional_values()` function in the R package *gratia* (version 0.11.1.9000, Simpson, 2024).

All figures were produced using the R packages *gratia* and *ggplot2* (version 3.5.2, Wickham, 2016).

## Results

*Lactation curves*

Figure 3a shows the three lactation curves estimated using Wood's model, a Tweedie GLM, and a Tweedie GAM. While all three models capture the general shape of the lactation data, the Tweedie GLM and, to a lesser extent, Wood's model, overestimate the peak yield, with the data exhibiting a broader period of peak fat content than is captured by either model. Wood's model and the Tweedie GLM also fail to capture the features of the mid–late lactation decline in fat content, other than the general decline itself. Conversely, the GAM, as anticipated, estimates a lactation curve that more faithfully tracks the observed data. The model response residuals ($y_i - \hat{y}_i$) for Wood's model (Figure 3b) and the Tweedie GLM (Figure 3c) show a significant amount of unmodelled signal, while the response residuals for the Tweedie GAM (Figure 3d) are much smaller and do not show a residual pattern. Despite using roughly twice as many degrees of freedom as the other models (Table 1), the Tweedie GAM was clearly favoured in terms of AIC and root mean squared error of the fitted values.

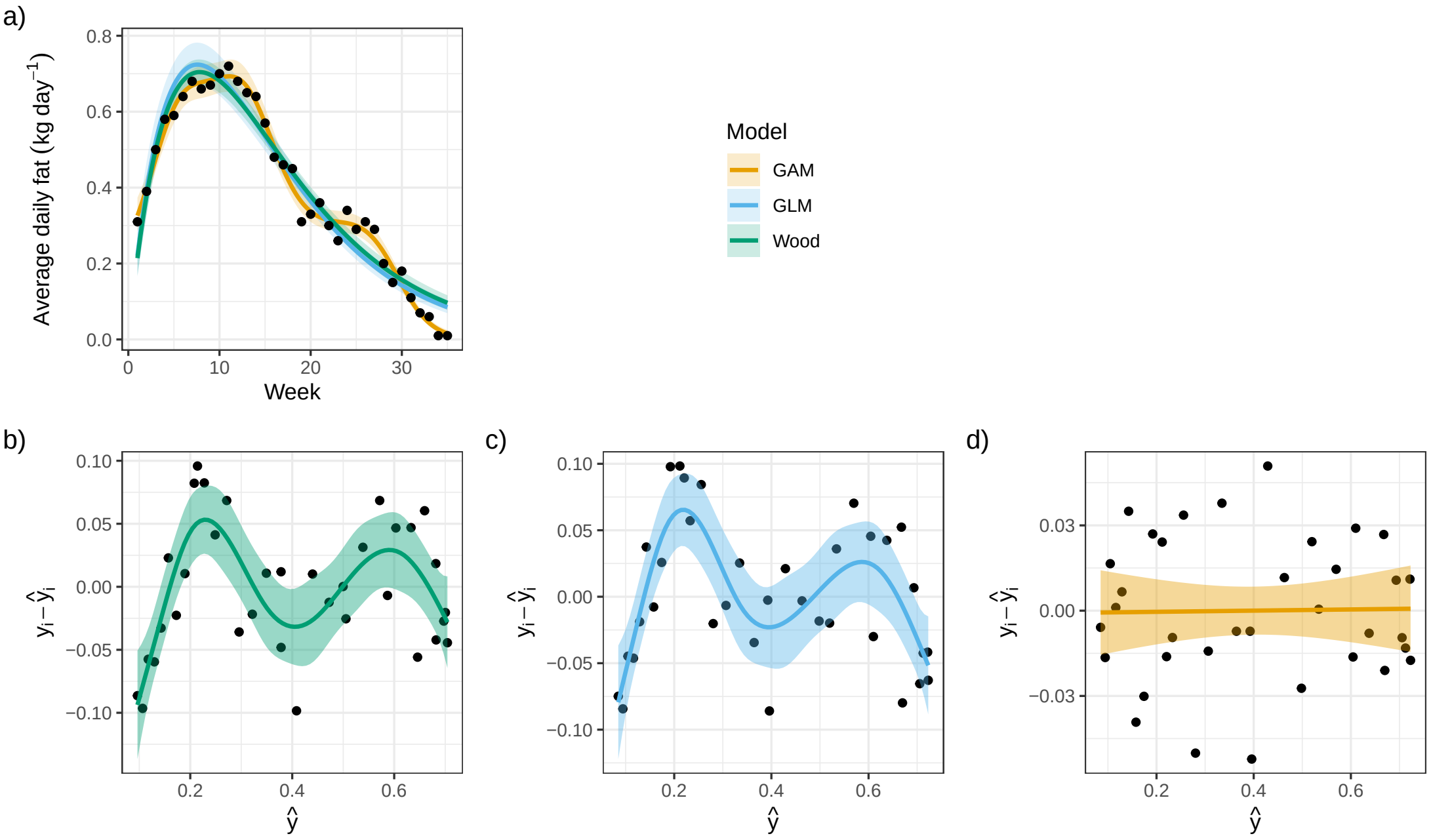

Fig. 3: Results of model fitting to the average daily fat content data from Henderson and McCulloch (1990). a) observed average daily fat content (points) and estimated lactation curves from Wood's (1967) model, a Tweedie GLM, and a Tweedie GAM (lines) with associated 95% confidence (Wood's model) or 95% credible intervals (GLM and GAM). Response residuals for Wood's model (b), Tweedie GLM (c), and Tweedie GAM (d), plus scatter plot smoothers (lines) and 95% credible intervals (shaded ribbons).

Figure 4 shows the estimated timing of peak fat content, the estimated fat content at the peak, and the first derivative of the estimated lactation curve at a point midway between the peak and the end of lactation, for the three estimated models. The peak in fat content is estimated to be significantly later in the GAM and the timing of the peak much more uncertain compared to both the fits from Wood's model and the GLM (Figure 4a). This reflects the observation that Wood's model and the GLM both over-estimate the fat content in the milk between weeks 5 and 10, and miss the broad period of high fat content up to week 14. Given the broad peak evident in the data, it is not unsurprising that the GAM-based estimate of the timing of the peak is much more uncertain than the other two estimates. The estimated peak fat content in the milk is broadly similar across the three models, with the GLM fit being slightly higher and more uncertain than either that for Wood's model or the GAM (Figure 4b). The rate of decay (first derivative) of the falling limb of the lactation curve is estimated to be substantially faster in both the GLM and Wood's model compared with the GAM estimate. This discrepancy is due to clear plateau in fat content between weeks 20 and 27 that is completely missed by the other two models. It should be noted that the posterior sampling method could equally be applied to yield an average derivative over the falling limb of the lactation curve if that was of greater focus.

Table 1: Comparison of models fitted to the lactation curve data set. Degrees of freedom and effective degrees of freedom ((E)DF) are shown for the Wood model and Tweedie GLM, and for the Tweedie GAM, respectively, showing the complexity in terms of (effective) numbers of parameters in each model. Akaike's An information criterion (AIC) values are shown for each model, alongside an estimate of the root mean squared error of the model fit estimated from the response residuals of each model.

| Model | (E)DF | AIC | RMSE |
|---|---|---|---|
| Wood | 4 | -100.6 | 0.3 |
| Tweedie GLM | 5 | -82.8 | 0.32 |
| Tweedie GAM | 8.004 | -133 | 0.15 |

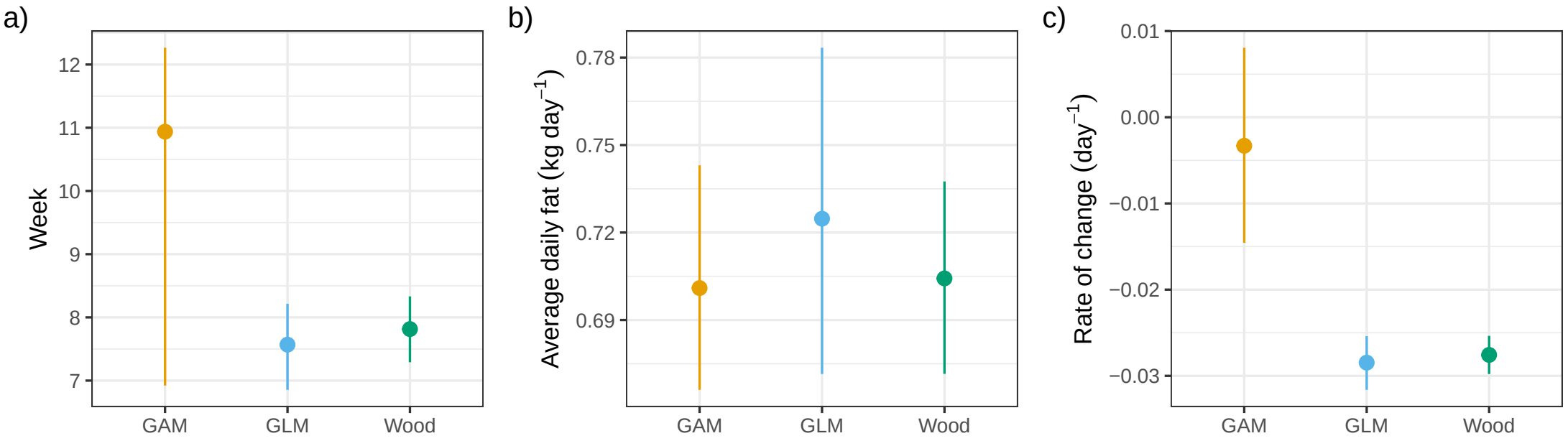

Fig. 4: Quantities of interest derived from Wood's model, a Tweedie GLM, and a Tweedie GAM fitted to the lactation data example: a) the estimated week of peak average daily fat content, b) the estimated average daily fat content at the peak, and c) the rate of change (first derivative) of the lactation curve estimated at a point that is midway between the peak fat content and the end of lactation. The points are the estimated values and the lines are a 95% uncertainty interval. The uncertainty interval is based on the 0.025 and 0.975 percentiles of the bootstrap distribution of model coefficient estimates (Wood's model) or of the posterior distribution (GLM and GAM).

For the timing of the peak and the rate of decay in the lactation curve, the estimates from the GLM and Wood's model are clearly overly optimistic in their uncertainty; the results in Figure 4a & c imply quite precisely estimated values and yet these models exhibit substantial biases in their estimated lactation curves, an issue that largely arises because they force a particular model to the data rather than allowing the data to inform the shape of the lactation curve.

*Pig growth*

The estimated growth curves for the 18 pigs in the pig growth example are shown in Figure 5, which were produced from model P2. Of the four models fitted, the two models that decomposed the growth effects into an average curve plus animal-specific deviations from the average curve, models P2 and P4, resulted in the most parsimonious fits from the point of view of AIC Table 2. This is due to these two model forms using many fewer degrees of freedom (~64) than either of models P1 (EDF = 83.54) or P3 (EDF = 90.57). Models P1 and P3 do not include the average curve and as a result expend many more degrees of freedom modelling the same general shape for each animal. Interestingly, for these data at least, allowing for a different smoothing parameter (P2) for each animals' deviation smooth seems to be preferred over using a single smoothing parameter (P4). In part, this is due to the somewhat idiosyncratic nature of each animal's growth curve in this data set.

Table 2: Comparison of models fitted to the pig weight data set. The model label is shown (see text for the specific model formulations). Effective degrees of freedom (EDF) represents model complexity in terms of the (effective) number of parameters in each model. Akaike's An information criterion (AIC) values, model deviance, and deviance explained as a proportion are also reported.

| Model | EDF | AIC | Deviance | Deviance expl. |
|---|---|---|---|---|
| P1 | 83.5 | 3839 | 2.07 | 0.964 |
| P2 | 64.4 | 3797 | 2.09 | 0.964 |
| P3 | 90.6 | 3824 | 2.05 | 0.964 |
| P4 | 64.8 | 3809 | 2.15 | 0.963 |

Table 3: Summary for model P2 fitted to the pig growth data set. The model term for comparison with the descriptions in the text and the label reported by the software are both shown for clarity. $k$ is the number of basis functions *per smooth*, EDF is the effective degrees of freedom, a measure of the complexity of each term, $\text{EDF}_{\text{Ref.}}$ is the reference degrees of freedom used in the tests, F is the test statistic and $p$ the $p$ value of the null hypothesis of a flat constant function or functions. The tests for terms in rows 1–3 are Wald-like tests with appropriate degrees of freedom. The tests of smooth terms (rows 2–3) are described in Wood (2013a).

| Model term | Label | $k$ | EDF | $\text{EDF}_{\text{Ref.}}$ | F | $p$ |
|---|---|---|---|---|---|---|
| $\beta_0$ | Intercept | NA | 1 | 1 | 1329 | $<0.001$ |
| $f(\texttt{day})$ | s(day) | 9 | 7.33 | 8.12 | 1303 | $<0.001$ |
| $f_{a(i)}(\texttt{day})$ | s(day,animal) | 10 | 56.05 | 63.31 | 10 | $<0.001$ |

Although the deviance explained is very high (~96%) for all models, much of this is due to use of random effects to model differences between animals, and should not be taken as a sign that the model can effectively perfectly predict the weight of a pig under similar conditions; it is clear from Figure 5 that there is much unmodelled variation around the estimated growth curves. One feature of the data that I do not address here is the clear variation among animals in the variance of the depth camera-based weight measurements; animals 5, 6, and 9 in particular, exhibit substantially greater variation than the other animals in the data set. I return to this point briefly later.

Table 3 provides an overview of the two terms in model P2. Although the average curve ($f(\texttt{day}_i)$) was allowed to use $K = 9$ basis functions, the wiggliness penalty has shrunk this back to 7.332 effective degrees of freedom (EDF). The animal-specific deviation smooths, $f_{\text{a}(i)}(\texttt{day}_i)$, were fully penalised and as such used $K = 10$ basis functions per animal. Here we clearly see the effect of the wiggliness penalty, which has resulted in a reduction from a potential 180 EDF for the set of animal-specific smooths to 56.054 EDF. The test of the null hypothesis for the average growth curve and the omnibus test for the pig-specific deviation smooths indicate both are statistically interesting effects.

Figure 6a shows the estimated daily growth rate on November 15$^{\text{th}}$, 2021 for each of the 18 pigs. Despite the fact that with GAMs we do not have a simple equation that can be used for estimation of biologically-relevant values using the model, posterior simulation is a simple and effective way to estimate values of interest, such as the derivatives of the response, and provide an uncertainty estimate. For three of the pigs in Figure 6a, no weight estimates were available after November 1$^{st}$, so the estimated growth curve for these pigs is not constrained by data on these animals, but instead is estimated using the shape of the average growth curve ($f(\texttt{day}_i)$). These estimates do include an animal-specific component, however, because we see quite differently-shaped posterior distributions for pig 17 compared with pigs 2 and 13, with the estimated growth curve of pig 17 being considerably more uncertain than either of the curves for the other two pigs. This is reflected in the much broader posterior distribution for pig 17 compared to those for pigs 2 and 13 Figure 6b.

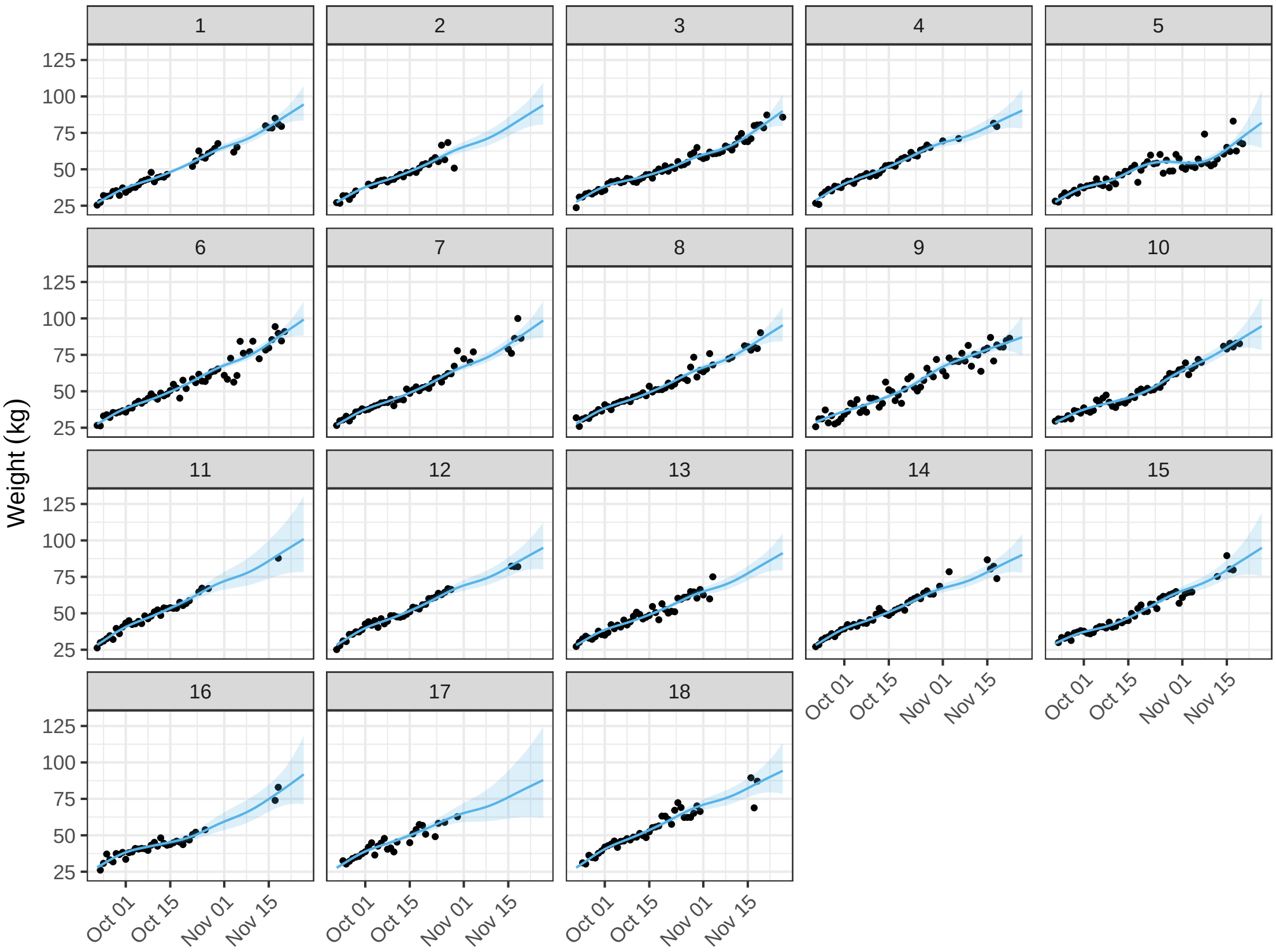

Fig. 5: Depth camera-based weight estimates from 18 commerical pigs. The data (black points) are the average of multiple measurements taken of each animal per day, 1 panel per pig. The panel labels indicate the pig shown. The estimated growth curve for each pig obtained using generalized additive model P2 is shown by the blue line in each panel. The blue shaded ribbon is the 95% Bayesian credible interval around the estimate curve.

*Quail hormone experiment*

As assessed by AIC, models containing treatment effects resulted in no improvement in the model fit (Table 4), a result that is consistent with the findings of Sarraude et al. (2020a). However, the results of model Q3 are reported below as this model is consistent with the *a priori* assumption that treatment effects might be present, possibly varying by sex, which underlay the original experiment. If we were to choose either of models Q1 or Q2 *on the basis of AIC*, we would be engaging in post-selection inference; having used the same data to both fit and test the model, any tests we would wish to conduct on the chosen model would yield anti-conservative $p$ values, and standard errors and credible intervals that would be too small or narrow respectively. Furthermore, choosing models Q1 or Q2 would in effect be claiming that some or all treatment effects were exactly equal to 0 in size; this is implausible and also inconsistent with the data we have observed. It is therefore better statistical practice to fit the model containing the effects hypothesised *a priori*, and report the estimated effect sizes of those effects.

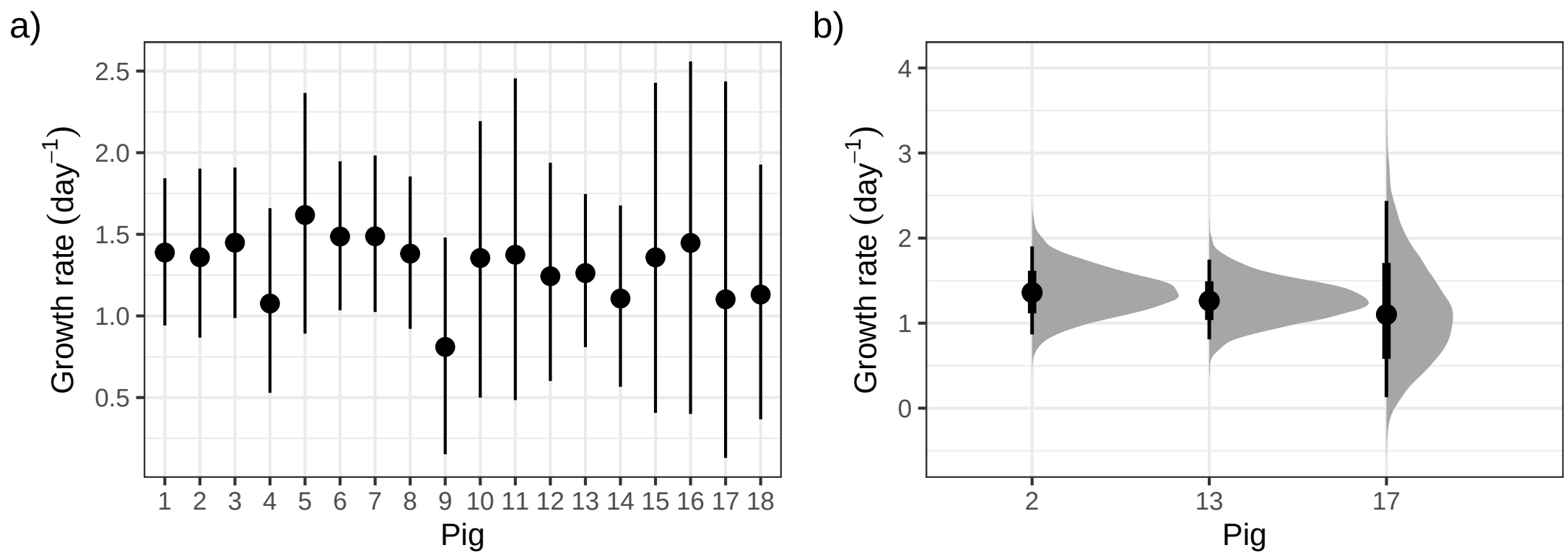

Fig. 6: a) Estimated daily growth rate on November 15$^{th}$, 2021 and 95% Bayesian credible interval for the 18 pigs in the pig growth example. b) Posterior distribution of daily growth rate on November 15$^{th}$, 2021, for three pigs (numbers 2, 13, and 17), for whom weight observations ceased before November 1$^{st}$, 2021. In b), the shaded region is the posterior distribution, the point, and thick and thin bars are the posterior median, and 66% and 95% posterior intervals respectively.

Table 4: Comparison of models fitted to the quail hormone experimental data set. The model label is shown (see text for the specific model formulations). Effective degrees of freedom (EDF) represents model complexity in terms of the (effective) number of parameters in each model. Akaike's An information criterion (AIC) values, model deviance, and deviance explained as a proportion are also reported.

| Model | EDF | AIC | Deviance | Deviance expl. |
|---|---|---|---|---|
| Q1 | 275 | 4458 | 12.2 | 0.998 |
| Q2 | 274.8 | 4459 | 12.4 | 0.998 |
| Q3 | 274.8 | 4459 | 12.5 | 0.998 |
| Q4 | 77.1 | 4987 | 30.4 | 0.995 |

Figure 7 shows the original data and the estimated growth curves for each individual quail that were obtained from model Q3. The need for quail-specific random smooths is clear enough from the data; model Q4 had the same model structure as that of Q3, except for the replacement of the quail-specific random smooths with quail-specific intercepts, which resulted in an increase in AIC of over 500 units (Table 4).

To focus on the estimated treatment effects, model Q3 was evaluated at 100 evenly-spaced values over the time covariate (the number of evaluation points was chosen to obtain a visually smooth representation of the estimated functions), plus all combinations of treatment level and sex. The effects of the quail-specific random smooths and the maternal random effects were excluded from these estimates. This results in estimated treatment effects for the average quail, which, strictly speaking should not be interpretted as population level effects due to the non-identity link function. The resulting estimated effects are shown in Figure 8. While there are clear differences in the growth curves based on sex, there appears to be little

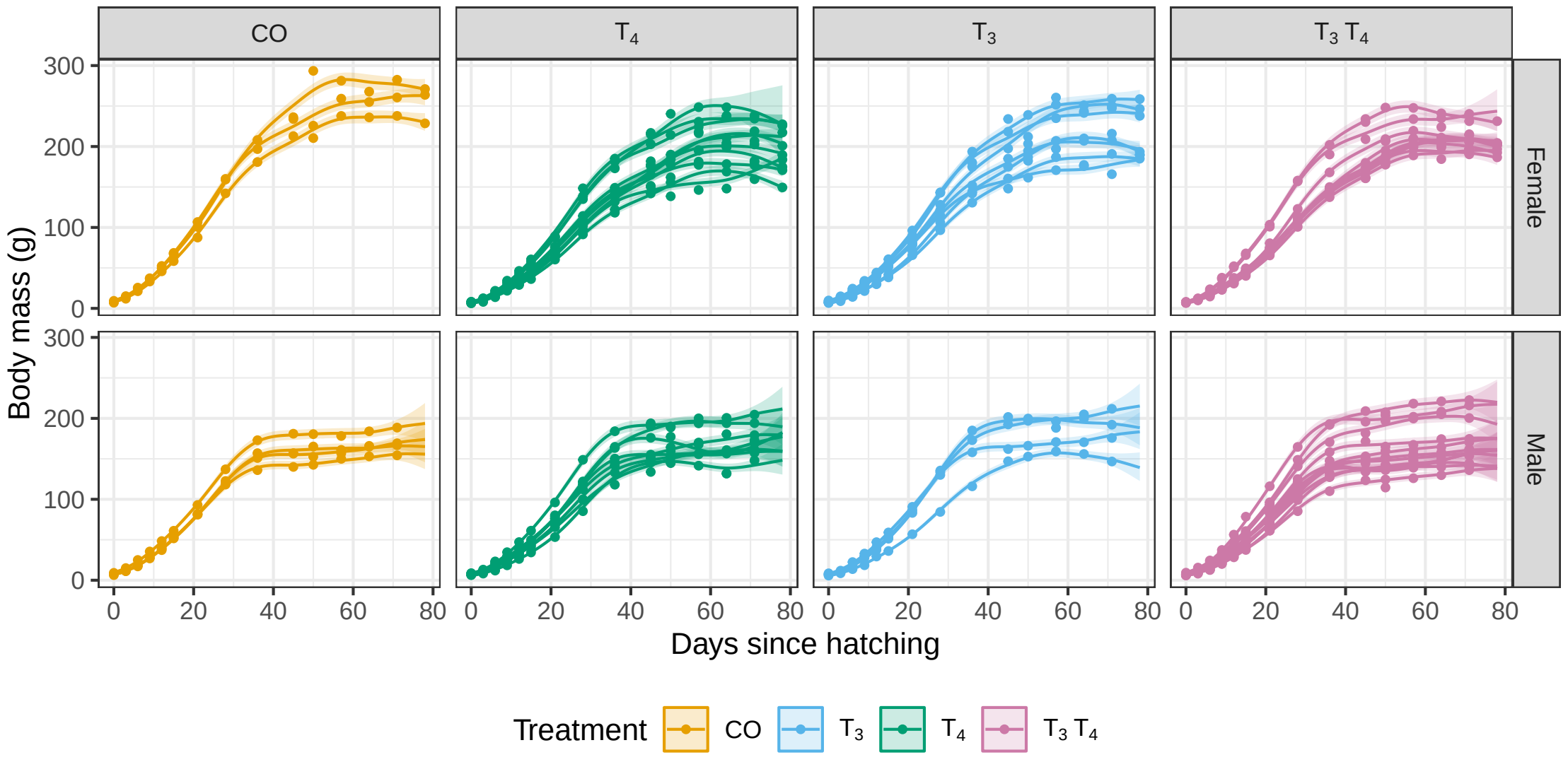

Fig. 7: Measured body mass (g) for 57 Japanese quail from an experiment on the effects of exposure to the maternal hormone $T_4$, its active metabolite $T_3$, both ($T_3T_4$) or a saline solution control (CO). The data (points) are shown along side the estimated growth curve for each quail obtained using generalized additive model Q3, which are shown by the coloured lines in each panel. The coloured shaded ribbon is the 95% Bayesian credible interval around the estimate curve. The data are faceted by treatment and the sex of bird.

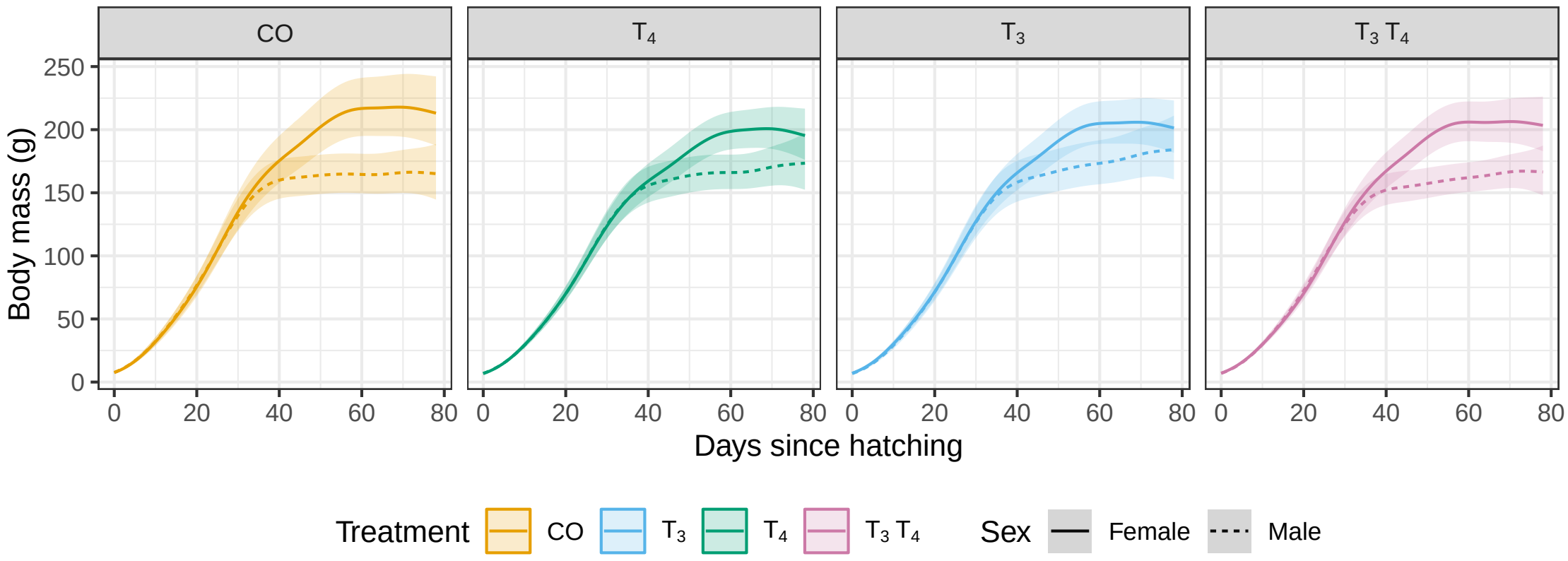

Fig. 8: Conditional value plots showing the growth rate for an average quail in each of the treatment groups by sex. The panels show the estimated curves for a particular treatment group; saline solution controls (CO), maternal hormone metabolite $T_3$, the maternal hormone $T_4$, and a combination of both $T_3T_4$. The estimated curve for male birds is shown by the solid line, and females the dashed line in each panel. The shaded band around each curve is a 95% Bayesian credible interval.

qualitative difference in the effects of treatment levels on quail growth.

The statistical summary of model Q3 is shown in Table 5. The omnibus tests of the treatment specific deviations (row 3; $f_{\text{treat(i)}}(\texttt{day}_i)$), and the treatment by sex deviations from the average curve (row 5; $f_{\text{treat}(i),\text{sex}(i)}(\texttt{day}_i)$), both have $p > 0.05$, which further reinforces the previously made observations that the effects of exposure to maternal thyroid hormone, if any, are small relative to the uncertainty in the model itself. The $p$ value for the omnibus test of the quail-specific deviation smooths ($f_{\text{egg}(i)}(\texttt{day}_i)$) is also greater than 0.05. This is somewhat surprising, given the magnitude of the difference in AIC that is observed when these quail-specific random smooths are replace by a simple random intercept term ($\Delta$AIC = 527.388). Despite not achieving "statistical significance", removing these terms on the basis of $p$ values would lead to invalid inference, which is to be avoided.

Table 5: Summary for model Q3 fitted to the quail hormone experimental data set. The model term for comparison with the descriptions in the text and the label reported by the software are both shown for clarity. $k$ is the number of basis functions *per smooth*, EDF is the effective degrees of freedom, a measure of the complexity of each term, $\text{EDF}_{\text{Ref.}}$ is the reference degrees of freedom used in the tests, F is the test statistic and $p$ the $p$ value of the null hypothesis of a flat constant function or functions. The tests for terms in rows 1–6 are Wald-like tests with appropriate degrees of freedom. The tests of smooth terms (rows 2–6) are described in Wood (2013a). The test of the random intercept term (row 7) is adjusted for testing a variance component whose null hypothesis lies on the boundary of the allowable parameter space (Wood, 2013b).

| Model term | Label | $k$ | EDF | $\text{EDF}_{\text{Ref.}}$ | F | $p$ |
|---|---|---|---|---|---|---|
| $\beta_0$ | Intercept | NA | 1 | 1 | 142.09 | <0.001 |
| $f(\texttt{day})$ | s(day) | 9 | 8.95 | 8.99 | 4797.2 | <0.001 |
| $f_{\text{treat(i)}}(\texttt{day}_i)$ | s(day,treat) | 10 | 6.01 | 6.01 | 1.66 | 0.129 |
| $f_{\text{sex(i)}}(\texttt{day}_i)$ | s(day,sex) | 10 | 7.32 | 8.25 | 14.31 | <0.001 |
| $f_{\text{treat}(i),\text{sex}(i)}(\texttt{day}_i)$ | s(day,treat,sex) | 10 | 9.73 | 10.71 | 1.56 | 0.111 |
| $f_{\text{egg}(i)}(\texttt{day}_i)$ | s(day,egg) | 6 | 224.06 | 326 | 67.7 | 0.267 |
| $\xi_{\text{mother}(i)}$ | s(mother) | 21 | 17.69 | 21 | 8.63 | <0.001 |

The largest contribution to the overall EDF of this model (EDF = 274.75) is from the quail-specific deviation smooths (EDF = 224.06), although because there are 57 individual animals and the random smooths include random intercepts to acocunt for differences in the average weight of each quail, this is unsurprising. Again, we see the effect of the wiggliness penalty, which has penalised the smooths back to EDF = 224.06 from a maximum EDF of 342.

To formally assess the differences among treatment effects, pairwise comparisons of treatment levels within sex were conducted for the slope of the average growth curve at day 20. The estimated differences in the slopes of the growth curves are shown in Table 6. For all comparisons, despite some observed differences in the growth rates at day 20 among the treatment levels, the 95% credible intervals include zero for all comparisons. Therefore, if these results reflect the broader population of quail, we should expect that any differences in growth rate due to maternal hormones are small, and largely indistinguishable from zero.

Finally, pairwise comparisons of quail weight at the end of the experiment (day 78) among treatment levels within sex were used to examine differences in estimated weight of quail due to exposure to maternal thyroid hormones. The results of these comparisons are shown in Table 7. The largest observed difference (~10.1 g) in weight of an average quail was between the $T_4$ hormone treatment and controls in female quail. This represents approximately a 5% decrease in the weight of an average female quail when exposed to the $T_4$ hormone compared with the untreated controls. Yet, given the uncertainty in the estimates of the model coefficients, the 95% credible interval includes 0 for this, and all other, comparisons.

## Author's Points of View

The results presented above demonstrate the utility of GAMs for modelling the kinds of data commonly encountered in research involving animals. GAMs are often viewed unfavourably as being subjective — requiring the user to specify how complex they want the estimated smooth functions to be — and data hungry, and more of a data visualisation tool than one suited to formal statistical inference. These views are largely out-dated with respect to the modern approach to GAMs described here. The subjectivity critique has been addressed through developments in automated smoothness selection. None of the data sets analysed here are especially large by today's standards, and where data sets are much larger, significant research effort has been (e.g., Li and Wood, 2020) and continues to be expended to adapt the algorithms to ever higher dimensional problems. The `bam()` function in the *mgcv* package, for example, can handle data on the order of millions of rows, with tens of thousands of model parameters, given sufficient availability of computer memory.

Table 6: Pairwise comparisons of treatment effects by sex on the slopes of the growth curves at day 20 for an average quail of the indicated sex in the pair of treatments shown. The Hypothesis column lists the specific pairwise comparison, Diff. is the estimated difference in the slope of the growth curves compared, SE its standard error. Z is the Wald test statistic, $p$ its $p$ value, and associated endpoints of a Bayesian 95% credible interval on the estimated difference of slopes.

| Sex | Hypothesis | Diff. | SE | Z | $p$ | 2.5% | 97.5% |
|---|---|---|---|---|---|---|---|
| Female | $T_4$ − Control | -0.3988 | 0.276 | -1.445 | 1 | -1.171 | 0.373 |
| Female | $T_3$ − Control | -0.2868 | 0.262 | -1.097 | 1 | -1.018 | 0.445 |
| Female | $T_3$ $T_4$ − Control | -0.3409 | 0.269 | -1.266 | 1 | -1.094 | 0.412 |
| Female | $T_3 - T_4$ | 0.112 | 0.207 | 0.541 | 1 | -0.467 | 0.691 |
| Female | $T_3$ $T_4 - T_4$ | 0.0579 | 0.257 | 0.225 | 1 | -0.66 | 0.775 |
| Female | $T_3$ $T_4 - T_3$ | -0.0541 | 0.218 | -0.248 | 1 | -0.664 | 0.556 |
| Male | $T_4$ − Control | -0.2884 | 0.251 | -1.148 | 1 | -0.991 | 0.414 |
| Male | $T_3$ − Control | -0.1443 | 0.273 | -0.529 | 1 | -0.907 | 0.618 |
| Male | $T_3$ $T_4$ − Control | -0.2033 | 0.233 | -0.871 | 1 | -0.856 | 0.449 |
| Male | $T_3 - T_4$ | 0.1441 | 0.252 | 0.572 | 1 | -0.56 | 0.848 |
| Male | $T_3$ $T_4 - T_4$ | 0.0851 | 0.274 | 0.31 | 1 | -0.682 | 0.852 |
| Male | $T_3$ $T_4 - T_3$ | -0.059 | 0.261 | -0.226 | 1 | -0.79 | 0.672 |

Table 7: Pairwise comparisons of treatment effects by sex on the estimated mean of the quail growth curves at day 78 for an average quail of the indicated sex in each pair of treatments. The Hypothesis column lists the specific pairwise comparison, Diff. is the estimated difference of mean body mass (g) of the growth curves compared, SE its standard error. Z is the Wald test statistic, $p$ its $p$ value, and associated endpoints of a Bayesian 95% credible interval on the estimated difference of means.

| Sex | Hypothesis | Diff. | SE | Z | $p$ | 2.5% | 97.5% |
|---|---|---|---|---|---|---|---|
| Female | $T_4$ − Control | -9.985 | 5.7 | -1.75 | 1 | -25.96 | 5.99 |
| Female | $T_3$ − Control | -6.279 | 5.82 | -1.079 | 1 | -22.57 | 10.01 |
| Female | $T_3T_4$ − Control | -5.529 | 5.59 | -0.989 | 1 | -21.18 | 10.12 |
| Female | $T_3 - T_4$ | 3.706 | 4.02 | 0.922 | 1 | -7.55 | 14.96 |
| Female | $T_3$ $T_4 - T_4$ | 4.456 | 3.63 | 1.227 | 1 | -5.72 | 14.63 |
| Female | $T_3T_4 - T_3$ | 0.75 | 4.34 | 0.173 | 1 | -11.4 | 12.9 |
| Male | $T_4$ − Control | -0.684 | 4.5 | -0.152 | 1 | -13.28 | 11.92 |
| Male | $T_3$ − Control | 2.558 | 5.37 | 0.476 | 1 | -12.49 | 17.6 |
| Male | $T_3T_4$ − Control | -2.921 | 4.02 | -0.727 | 1 | -14.17 | 8.33 |
| Male | $T_3 - T_4$ | 3.242 | 4.68 | 0.692 | 1 | -9.87 | 16.36 |
| Male | $T_3$ $T_4 - T_4$ | -2.238 | 3.71 | -0.604 | 1 | -12.61 | 8.14 |
| Male | $T_3T_4 - T_3$ | -5.479 | 4.81 | -1.14 | 1 | -18.94 | 7.98 |

As demonstrated in the quail maternal hormone example, formal statistical inference is entirely feasible. If we ignore the selection of smoothness parameters, the models described here are little more than generalized linear (mixed) models (GL(M)Ms) once the basis expansions of covariates have been performed. Modern software tools like the *marginaleffects* package for R allow a consistent interface to statistical inference across many disparate statistical models, and GAMs are no different in this regard. Posterior sampling can also be applied to GAMs to provide point estimates and uncertainties on any quantity of interest that can be computed using only the model predictions. This sampling-based approach to inference is incredibly powerful and extremely flexible and complements more standard inference such as the comparisons of contrasts used in the quail example.

The distinct advantage of GAMs over GL(M)Ms is their ability to learn the shapes of nonlinear relationships between covariates and the response from the data themselves. This relieves the analyst from having to force data to fit a particular theoretical model, unless they have a good justification via domain knowledge to use that model. GAMs also avoid the model selection problems and erratic estimates inherent to modelling with polynomial basis expansions.

The main disadvantage of GAMs is that they are more complex to fit than GL(M)Ms and the analyst has some additional data modelling choices to make. The main choice is the number of basis functions, $K$, that should be used by each smooth in the model. The general advice here is for the analyst to imagine the largest amount of wiggliness that they would expect and then set $K$ a little larger than this. Of course, the novice user of GAMs will lack the experience required to do this easily, but for the sorts of data problems exemplified above, we would not expect highly complex smooth functions, and the default of $K = 10$ for univariate smooths in *mgcv* is usually sufficient for many problems. The key requirement is that the initial number of basis functions used should be large enough such that the span of functions representable with that basis will contain the true but unknown function or a close approximation to it. The basis dimension must be checked of course, and this adds another step to the model appraisal or checking procedure. With *mgcv* for example, the `k.check()` function provides a test for sufficiency of the basis size used to fit the model (Pya and Wood, 2016).

A further disadvantage is that statistical inference is somewhat more approximate than with GL(M)Ms. While GAMs share with GL(M)Ms the property of having asymptotically correct $p$ values, the $p$ values for smooths are more approximate than for terms in a GL(M)M because the current theory on which these tests are based does not account for the selection of smoothing parameters; for the purposes of the tests, the smoothing parameters are treated as being fixed and known, but instead they are estimated from the data (Wood, 2013a). While there has been some progress in adapting the theory to include this additional source of uncertainty (e.g., Wood et al., 2016), as yet this has not been used to correct the $p$ values of tests of smooths. That said, we should avoid making modelling decisions on the basis of $p$ values, to this additional approximation should not pose a problem in practice.

Finally, GAMs can require more substantial amounts of computing resources to fit than GL(M)Ms once data sets get above tens of thousand observations or where models include several or complex random effect terms, using the algorithms provided by *mgcv*. The main requirement is computer memory, although the `bam()` function can help with this using algorithmic improvements from Li and Wood (2020). The examples above were all run on an Apple M1 Pro MacBook Pro with 32GB of RAM, and the entire analysis including the generations of figures takes only a few minutes.

GAMs are a very broad and general class of models, and many extensions exist in the literature. In the case of the pig growth data, several animals exhibited visibly more variation in their weight measurements than the majority of the animals in the data. Such heteroscedasticity (non-constant variance) can be modelled using distributional GAMs (or Generalized additive models for location, scale, shape or GAMLSS) (e.g. Rigby and Stasinopoulos, 2005; Kneib et al., 2021; Klein, 2024), which include linear predictors for all of the parameters of a distribution, or centile or quantile models (e.g., Nakamura et al., 2022). In the examples above, the additional parameters (scale in the case of the Gamma, or the Tweedie power parameter) were estimated as constants for the entire data set. A distributional GAM would allow those parameters to potentially vary with individual animals, or as as smooth functions of the covariates, just as was done for the mean of the distribution in this study.

In conclusion, GAMs are a modern, flexible, and highly usable statistical model that is amenable to many research problems in animal science, and deserve a place in the statistical toolbox.

### Ethics approval

Not applicable. See the original sources of the data used for ethical approval.

### Declaration of Generative AI and AI-assisted technologies in the writing process

The author did not use any artificial intelligence technologies in the writing process.

### Author ORCIDs

**Gavin L. Simpson**: 0000-0002-9084-8413

### Author contributions

GLS: Conceptualization, Methodology, Software, Formal analysis, Writing, Visualization.

### Declaration of interest

None.

## Acknowledgements

The author would to like to express their appreciation to colleagues at the Department of Animal and Veterinary Sciences, Aarhus University for making the pig growth data available, and in particular to Dr. Mona Larsen for supplying the data and arranging with their coauthors to enable the subset analysed in this manuscript to be made available as open data.

## Financial support statement

This work was supported by an Aarhus Universitets Forskningsfond (Aarhus University Research Foundation; AUFF) starting grant awarded to the author.

## References

Arel-Bundock, V., Greifer, N., Heiss, A., 2024. How to interpret statistical models using marginaleffects for *R* and *Python*. J. Stat. Softw. 111, 1–32. URL: https://www.jstatsoft.org/index.php/jss/article/view/v111i09, doi:10.18637/jss.v111.i09.

Baty, F., Ritz, C., Charles, S., Brutsche, M., Flandrois, J.P., Delignette-Muller, M.L., 2015. A toolbox for nonlinear regression in R: The package nlstools. Journal of Statistical Software 66, 1–21. doi:10.18637/jss.v066.i05.

Benjamini, Y., Yekutieli, D., 2001. The control of the false discovery rate in multiple testing under dependency. Ann. Stat. 29, 1165–1188. URL: http://projecteuclid.org/download/pdf_1/euclid.aos/1013699998.

Benni, S., Pastell, M., Bonora, F., Tassinari, P., Torreggiani, D., 2020. A generalised additive model to characterise dairy cows' responses to heat stress. Animal 14, 418–424. URL: http://dx.doi.org/10.1017/S1751731119001721, doi:10.1017/S1751731119001721.

Brito, L.F., Gomes da Silva, F., Rojas de Oliveira, H., Souza, N., Caetano, G., Costa, E.V., Romeiro de Oliveira Menezes, G., Puerro de Melo, A.L., Teixeira Rodrigues, M., de Almeida Torres, R., 2017. Modelling lactation curves of dairy goats by fitting random regression models using legendre polynomials or b-splines. Can. J. Anim. Sci. URL: http://dx.doi.org/10.1139/cjas-2017-0019, doi:10.1139/cjas-2017-0019.

Bus, J.D., Franchi, G.A., Boumans, I.J.M.M., Te Beest, D.E., Webb, L.E., Jensen, M.B., Pedersen, L.J., Bokkers, E.A.M., 2025. Short-term associations between ambient ammonia concentrations and growing-finishing pig performance and health. Prev. Vet. Med. 242, 106555. URL: http://dx.doi.org/10.1016/j.prevetmed.2025.106555, doi:10.1016/j.prevetmed.2025.106555.

Dijkstra, J., Lopez, S., Bannink, A., Dhanoa, M.S., Kebreab, E., Odongo, N.E., Fathi Nasri, M.H., Behera, U.K., Hernandez-Ferrer, D., France, J., 2010. Evaluation of a mechanistic lactation model using cow, goat and sheep data. J. Agric. Sci. 148, 249–262. URL: http://dx.doi.org/10.1017/S0021859609990578, doi:10.1017/s0021859609990578.

Eilers, P.H.C., Marx, B.D., 1996. Flexible smoothing with B-splines and penalties. Stat. Sci. 11, 89–121. URL: https://projecteuclid.org/journals/statistical-science/volume-11/issue-2/Flexible-smoothing-with-B-splines-and-penalties/10.1214/ss/1038425655.short, doi:10.1214/ss/1038425655.

Franchi, G.A., Bus, J.D., Boumans, I.J.M.M., Bokkers, E.A.M., Jensen, M.B., Pedersen, L.J., 2023. Estimating body weight in conventional growing pigs using a depth camera. Smart Agric. Technol. 3, 100117. URL: http://dx.doi.org/10.1016/j.atech.2022.100117, doi:10.1016/j.atech.2022.100117.

Hastie, T.J., Tibshirani, R.J., 1990. Generalized Additive Models. Chapman & Hall / CRC. URL: https://market.android.com/details?id=book-qa29r1Ze1coC.

Henderson, H.V., McCulloch, C., 1990. Transform or link. URL: https://ecommons.cornell.edu/bitstream/1813/31620/1/BU-1049-MA.pdf.

Hirst, W.M., Murray, R.D., Ward, W.R., French, N.P., 2002. Generalised additive models and hierarchical logistic regression of lameness in dairy cows. Prev. Vet. Med. 55, 37–46. URL: http://dx.doi.org/10.1016/S0167-5877(02)00058-2, doi:10.1016/s0167-5877(02)00058-2.

Huang, C.H., Furukawa, K., Kusaba, N., 2023. Estimating the nonlinear interaction between somatic cell score and differential somatic cell count on milk production by parity using generalized additive models. J. Dairy Sci. URL: http://dx.doi.org/10.3168/jds.2022-22958, doi:10.3168/jds.2022-22958.

Klein, N., 2024. Distributional regression for data analysis. Annu. Rev. Stat. Appl. 11. URL: https://www.annualreviews.org/content/journals/10.1146/annurev-statistics-040722-053607, doi:10.1146/annurev-statistics-040722-053607.

Kneib, T., Silbersdorff, A., Säfken, B., 2021. Rage against the mean – a review of distributional regression approaches. Econometrics and Statistics URL: https://www.sciencedirect.com/science/article/pii/S2452306221000824, doi:10.1016/j.ecosta.2021.07.006.

Li, Z., Wood, S.N., 2020. Faster model matrix crossproducts for large generalized linear models with discretized covariates. Stat. Comput. 30, 19–25. URL: https://doi.org/10.1007/s11222-019-09864-2, doi:10.1007/s11222-019-09864-2.

van Lingen, H.J., Fadel, J.G., Kebreab, E., Bannink, A., Dijkstra, J., van Gastelen, S., 2023. Smoothing spline assessment of the accuracy of enteric hydrogen and methane production measurements from dairy cattle using various sampling schemes. J. Dairy Sci. 106, 6834–6848. URL: http://dx.doi.org/10.3168/jds.2022-23207, doi:10.3168/jds.2022-23207.

Macciotta, N.P.P., Dimauro, C., Rassu, S.P.G., Steri, R., Pulina, G., 2011. The mathematical description of lactation curves in dairy cattle. Ital. J. Anim. Sci. 10, e51. URL: http://dx.doi.org/10.4081/ijas.2011.e51, doi:10.4081/ijas.2011.e51.

Macciotta, N.P.P., Miglior, F., Dimauro, C., Schaeffer, L.R., 2010. Comparison of parametric, orthogonal, and spline functions to model individual lactation curves for milk yield in canadian holsteins. Ital. J. Anim. Sci. 9, e87. URL: https://doi.org/10.4081/ijas.2010.e87, doi:10.4081/ijas.2010.e87.

Miller, D.L., 2025. Bayesian views of generalized additive modelling. Methods Ecol. Evol. URL: https://onlinelibrary.wiley.com/doi/abs/10.1111/2041-210X.14498, doi:10.1111/2041-210x.14498.

Morota, G., Cheng, H., Cook, D., Tanaka, E., 2021. Asas-nanp symposium: prospects for interactive and dynamic graphics in the era of data-rich animal science1. J. Anim. Sci. 99. URL: http://dx.doi.org/10.1093/jas/skaa402, doi:10.1093/jas/skaa402.

Nagel-Alne, G.E., Krontveit, R., Bohlin, J., Valle, P.S., Skjerve, E., Sølverød, L.S., 2014. The norwegian healthier goats program–modeling lactation curves using a multilevel cubic spline regression model. J. Dairy Sci. 97, 4166–4173. URL: http://dx.doi.org/10.3168/jds.2013-7228, doi:10.3168/jds.2013-7228.

Nakamura, L.R., Ramires, T.G., Righetto, A.J., Pescim, R.R., Roquim, F.V., Savian, T.V., Stasinopoulos, D.M., 2022. Cattle reference growth curves based on centile estimation: A gamlss approach. Comput. Electron. Agric. 192, 106572. URL: http://dx.doi.org/10.1016/j.compag.2021.106572, doi:10.1016/j.compag.2021.106572.

Oliveira, J.G.d., Sant'Anna, D.F.D., Lourenço, M.C., Tavares, D.S.T., Rodrigues, M.T., Tedeschi, L.O., Vieira, R.A.M., 2020. The geometry of the lactation curve based on wood's equation: a two-step prediction. Rev. Bras. Zootec. 49, –. URL: http://dx.doi.org/10.37496/rbz4920200023, doi:10.37496/rbz4920200023.

Pedersen, E.J., Miller, D.L., Simpson, G.L., Ross, N., 2019. Hierarchical generalized additive models in ecology: an introduction with mgcv. PeerJ 7, e6876. URL: http://dx.doi.org/10.7717/peerj.6876, doi:10.7717/peerj.6876.

Pya, N., Wood, S.N., 2016. A note on basis dimension selection in generalized additive modelling. arXiv [stat.ME] URL: http://arxiv.org/abs/1602.06696, arXiv:1602.06696.

R Core Team, 2025. R: A Language and Environment for Statistical Computing. R Foundation for Statistical Computing. Vienna, Austria. URL: https://www.R-project.org/.

Reiss, P.T., Ogden, R.T., 2009. Smoothing parameter selection for a class of semiparametric linear models. J. R. Stat. Soc. Series B Stat. Methodol. 71, 505–523. URL: http://doi.wiley.com/10.1111/j.1467-9868.2008.00695.x, doi:10.1111/j.1467-9868.2008.00695.x.

Rigby, R.A., Stasinopoulos, D.M., 2005. Generalized additive models for location, scale and shape. J. R. Stat. Soc. Ser. C Appl. Stat. 54, 507–554. URL: http://dx.doi.org/10.1111/j.1467-9876.2005.00510.x, doi:10.1111/j.1467-9876.2005.00510.x.

Sarraude, T., Hsu, B.Y., Groothuis, T., Ruuskanen, S., 2020a. Dataset of prenatal thyroid hormones manipulation in Japanese quails. URL: https://zenodo.org/record/3741711, doi:10.5281/zenodo.3741711.

Sarraude, T., Hsu, B.Y., Groothuis, T., Ruuskanen, S., 2020b. Testing the short-and long-term effects of elevated prenatal exposure to different forms of thyroid hormones. PeerJ 8, e10175. URL: http://dx.doi.org/10.7717/peerj.10175, doi:10.7717/peerj.10175.

Silvestre, A.M., Petim-Batista, F., Colaço, J., 2006. The accuracy of seven mathematical functions in modeling dairy cattle lactation curves based on test-day records from varying sample schemes. J. Dairy Sci. 89, 1813–1821. URL: http://www.journalofdairyscience.org/article/S0022030206722500/abstract, doi:10.3168/jds.S0022-0302(06)72250-0.

Simpson, G.L., 2024. gratia: An R package for exploring generalized additive models. J. Open Source Softw. 9, 6962. URL: https://joss.theoj.org/papers/10.21105/joss.06962, doi:10.21105/joss.06962.

Smyth, G., 1996. Regression modelling quantity data exact zeros, in: Wilson, R.J., Osaki, S., Murthy, D.N.P. (Eds.), Proceedings second Australia-Japan workshop stochastic models engineering, Technology Management Centre, The University of QueenslandTechnology Management Centre, The University of Queensland. pp. 572–580.

White, I.M., Thompson, R., Brotherstone, S., 1999. Genetic and environmental smoothing of lactation curves with cubic splines. J. Dairy Sci. 82, 632–638. URL: http://dx.doi.org/10.3168/jds.S0022-0302(99)75277-X, doi:10.3168/jds.S0022-0302(99)75277-X.

Wickham, H., 2016. ggplot2: Elegant Graphics for Data Analysis. Use R!, Springer International Publishing. URL: https://link.springer.com/book/10.1007/978-3-319-24277-4, doi:10.1007/978-3-319-24277-4.

Wood, P.D.P., 1967. Algebraic model of the lactation curve in cattle. Nature 216, 164–165. URL: http://dx.doi.org/10.1038/216164a0, doi:10.1038/216164a0.

Wood, S.N., 2003. Thin plate regression splines. J. R. Stat. Soc. Series B Stat. Methodol. 65, 95–114. URL: http://dx.doi.org/10.1111/1467-9868.00374, doi:10.1111/1467-9868.00374.

Wood, S.N., 2011. Fast stable restricted maximum likelihood and marginal likelihood estimation of semiparametric generalized linear models. J. R. Stat. Soc. Series B Stat. Methodol. 73, 3–36. URL: http://dx.doi.org/10.1111/j.1467-9868.2010.00749.x, doi:10.1111/j.1467-9868.2010.00749.x.

Wood, S.N., 2013a. On p-values for smooth components of an extended generalized additive model. Biometrika 100, 221–228. URL: http://biomet.oxfordjournals.org/content/100/1/221.abstract, doi:10.1093/biomet/ass048.

Wood, S.N., 2013b. A simple test for random effects in regression models. Biometrika 100, 1005–1010. URL: http://biomet.oxfordjournals.org/content/100/4/1005.abstract, doi:10.1093/biomet/ast038.

Wood, S.N., 2017. Generalized Additive Models: An Introduction with R, Second Edition. CRC Press. URL: https://market.android.com/details?id=book-JTkkDwAAQBAJ.

Wood, S.N., 2025. mgcv: Mixed GAM computation vehicle with automatic smoothness estimation. URL: http://dx.doi.org/10.32614/cran.package.mgcv, doi:10.32614/cran.package.mgcv.

Wood, S.N., Pya, N., Säfken, B., 2016. Smoothing parameter and model selection for general smooth models. J. Am. Stat. Assoc. 111, 1548–1563. URL: https://doi.org/10.1080/01621459.2016.1180986, doi:10.1080/01621459.2016.1180986, arXiv:http://dx.doi.org/10.1080/01621459.2016.1180986. doi: 10.1080/01621459.2016.1180986.

Yano, M., Shimadzu, H., Endo, T., 2014. Modelling temperature effects on milk production: a study on holstein cows at a japanese farm. SpringerPlus 3, 129. URL: http://dx.doi.org/10.1186/2193-1801-3-129, doi:10.1186/2193-1801-3-129.